\documentclass[iop]{emulateapj}

\begin{document}

\title{The structure and dynamics of molecular gas in planet-forming zones: A CRIRES spectro-astrometric survey}

\author{Klaus M. Pontoppidan\altaffilmark{1}}
\author{Geoffrey A. Blake\altaffilmark{2}}
\author{Alain Smette\altaffilmark{3}}

\altaffiltext{1}{Space Telescope Science Institute, Baltimore, MD 21218; pontoppi@stsci.edu}
\altaffiltext{2}{California Institute of Technology, Division of Geological and Planetary Sciences, 
MS 150-21, Pasadena, CA 91125}
\altaffiltext{3}{European Southern Observatory, Casilla 19001, Santiago 19, Chile}

\begin{abstract}
We present a spectro-astrometric survey of molecular gas in the inner regions of 16 protoplanetary disks using CRIRES, the high
resolution infrared imaging spectrometer on the Very Large Telescope. Spectro-astrometry with CRIRES measures the spatial extent of line emission to sub-milliarcsecond
precision, or $<0.2$\,AU at the distance of the observed targets. The sample consists of gas-rich disks surrounding stars with spectral types ranging from K to A.  
The properties of the spectro-astrometric signals divide the sources into two distinct phenomenological classes: one that shows clear Keplerian astrometric spectra, and one in which the astrometric signatures are dominated by gas with strong non-Keplerian (radial) motions. Similarly to the near-infrared continuum emission, as determined by interferometry, we find that the size of the CO line emitting region in the Keplerian sources obeys a size-luminosity relation as $R_{\rm CO}\propto L_*^{0.5}$. The non-Keplerian spectro-astrometric signatures are likely indicative of the presence of wide-angle disk winds. The central feature of the winds is a strong sub-Keplerian velocity field due to conservation of 
angular momentum as the wind pressure drives the gas outwards. We construct a parametrized 2-dimensional disk+wind model that reproduces the observed characteristics the observed CO spectra and astrometry. The modeled winds indicate mass-loss rates of $\gtrsim 10^{-10}-10^{-8}\,M_{\odot}\,\rm yr^{-1}$. We suggest a unifying model in which all disks have slow molecular winds, but where the magnitude of the mass-loss rate determines the degree to which the mid-infrared molecular lines are dominated by the wind relative to the Keplerian disk surface.
\end{abstract}


\section{Introduction}

Planets are believed to predominantly form in the inner regions of circumstellar disks surrounding young stars, the
so-called planet-forming zone (PFZ). Consequently, searches for embryonic planetary systems are expected to be most fruitful if they are focused
on the inner regions of planet-forming disks, whether using direct detection methods or indirect approaches such as imaging signatures of planet-disk interactions.
While typical gas-rich disks around young stars may extend to 100 AU, or more in a few extreme cases, the outer reaches are too tenuous to
form planetary cores within the lifetime of the disk, unless the disk is massive enough to become gravitationally unstable \citep{Boss97,Marois08}.
A few mature planets or substellar objects have been imaged at large radii around A stars, but the majority of the known planetary systems
are believed to have formed within 20 AU \citep{Pollack96}, followed by radial migration inwards due to planet-disk interactions {\it during the gas-rich phase of the disk} \citep{Alibert04}, 
consistent with the currently known distribution of exo-planets detected with the radial velocity and transit methods. 

\subsection{An observational challenge}
A primary difficulty in studying planet formation {\it in progress} is the small angular sizes of the PFZs. The nearest protoplanetary disks are located, 
with a few notable exceptions such as the TW~Hya association and a few Herbig Ae stars, at distances in excess of 100 pc. Resolved images of typical PFZs must be obtained
at a spatial resolution better than 0\farcs1. Further compounding the problem is that many of the main tracers of PFZs are found in the infrared
wavelength range (2-200\,$\mu$m) because the relevant gas temperatures are in excess of $100$\,K -- making ground-based observations challenging. 
Generally, specialized instrumentation is needed to obtain the 
requisite spatial resolution. In recent years, ground-breaking progress has been made in infrared interferometry of the innermost regions of
protoplanetary disks from 1-2\,$\mu$m \citep[e.g.,][]{Millan-Gabet99, Eisner05} and near 10\,$\mu$m \citep{vanBoekel05}. However, interferometric observations of gas
have thus far been limited to single band or relatively low spectral resolving power and to young A and B stars. 
This has, for the most part, limited interferometry to dust, gas continuum and hydrogen recombination lines \citep{Tatulli07,Tatulli08,Isella08,Eisner09}. 

Our understanding of the structure and dynamics of inner protoplanetary disks and PFZs remains limited. While there appears to be a common end result of disk evolution -- the formation of a planetary system, complete with gas giants and perhaps smaller rocky planets, as well as, presumably, planetesimals, comets and zodiacal dust -- the pathway is unclear. 
It is known that disks exist and that they carry most of the angular momentum
of a young stellar system, as evidenced by resolved sub-millimeter imaging of their outer regions \citep{Koerner93, Mannings97, Qi08}, and matching the momentum
distribution of the solar system and other planetary systems. Their spectral energy distributions show us that the disks are gas-rich throughout; if they were not, 
the disks would be flat, not flared with essentially pressure-supported scale heights \citep{Kenyon87}. 

The presence of accretion shocks and jets tell us that the disks are actively accreting requiring the 
gas to be sufficiently viscious to allow angular momentum transport \citep{Koenigl91}. Limited disk lifetimes of about 6 Myr have been inferred, albeit with significant variation, demonstrating the need
for efficient dispersal mechanisms \citep{Haisch01}. Disks do not extend all the way inward to the stellar surface, but exhibit a complex structure
due to the sublimation of dust at high temperatures leading to rapid opacity gradients \citep{Natta01}, and a coupling of the disk to the stellar magnetic field facilitates
both accretion and mass loss \citep{Koenigl91, Shu94}. 

Some disks appear to have large excavated inner regions of low dust opacity \citep{Strom89}, compared to the outer disk, but it appears that not
all disks go through such a stage \citep{Muzerolle10}, and it is still an open question whether this is directly related to the formation
of planetesimals or even giant planets, or whether another dispersal mechanism is at play \citep{Alexander09}. Clearly, observing what the gas actually does in the inner disk 
impacts our understanding of all these disk properties and enables us to estimate their relative importance for the evolution of
the disk. Ultimately, the motion of the inner disk gas may reveal the presence of accreting protoplanets \citep{Regaly10}. 

\subsection{This paper}
\label{this_paper}
We present a spectro-astrometric imaging survey of molecular gas in the PFZs of disks around a sample of solar type stars using
CRIRES on the European Very Large Telescope \citep{Kaufl04}. The primary goal is to directly determine
the basic distribution and kinematics of the gas and to relate this to the process of planet formation and inner disk evolution. For instance, 
in a purely passive, non-accreting, disk it may be expected that the gas orbits at essentially Keplerian speeds, dictated only by the mass of the central star, neglecting minor corrections for the mass of the disk itself ($dV/V_{\rm Kepler}\lesssim 10^{-2}$) and for the radial pressure gradient in the disk \citep[$dV/V_{\rm Kepler}\sim 10^{-3}$][]{Cuzzi93}. The presence of double-peaked emission line profiles from protoplanetary disks can indeed be explained by gas in Keplerian orbits \citep{Carr93, Blake04, Pontoppidan08}. However, disks are in general not passive as they accrete and eventually dissipate, so departures from pure Keplerian motion are expected at some level, and is indicated in FU Ori stars for the CO bandhead at 2.3\,$\mu$m \citep{Calvet91, Martin97, Hartmann04}.  Do mid-infrared molecular lines -- in fact -- trace material strictly in Keplerian motion, or are there significant non-Keplerian components present, and, if so, are they dominated by infall or outflow motions? What disk radii are traced by rovibrational CO lines; the inner edge of the disk at 0.01-0.1\,AU, the terrestrial planet region at 0.1-1.0\,AU or the giant planet region at 1.0-10\,AU? What is the origin of the absorption components seen in CO rovibrational spectra -- absorption from edge-on disks, foreground clouds, remnant envelopes or disk winds? 

Our spectro-astrometric survey was primarily focused on what is probably the best tracer of molecular gas in the inner disk, or at least the
most easily observable -- the fundamental rovibrational band of CO centered at 4.67\,$\mu$m. These CO lines
are traditionally thought to trace warm gas in disks at $\sim$$1\,$AU, based on typical line widths and excitation temperatures 
\citep{Najita03,Blake04}, and are invariably bright in nearly all classical T Tauri and Herbig Ae stars. One of the most basic outcomes
of this survey is the direct measurements of the size of the CO line emitting regions.

We demonstrate that astrometric signals in CO were detected for all sources on AU-scales, but with varying amplitude and with an intriguing range of structure:
The CO spectra are divided into three rough phenomenological classes, based on the line shape in combination with the 
shape of astrometric spectra: Keplerian disks characterized by double-peaked line-profiles, single-peaked line sources with broad wings, and self-absorbed sources. 

This paper is arranged as follows: In \S\ref{observations} the survey and data reduction are described, including a
detailed discussion of the capabilities of the spectro-astrometric mode of CRIRES for super-resolution imaging. 
In \S\ref{Kepler_sources}), we discuss the results of fitting simple Keplerian disk models to the data. 
The central issue is that many CO line and astrometric spectra {\it cannot} be explained by Keplerian velocity fields. 
We introduce a non-Keplerian class of emission lines in \S\ref{Non_Kepler} and explain why a purely Keplerian model fails. 
In \S\ref{wind_model}, we develop a 2-dimensional model that adds a disk wind to a regular Keplerian, flared disk, and demonstrate
that this provides a framework for matching all CO line and astrometric spectra from classical T Tauri stars under specific circumstances, namely if the wind is slow
and uncollimated. We suggest that there is a smooth transition from lines dominated by Keplerian motions to wind-dominated lines, possibly
scaling with the mass-loss/accretion rates. In \S\ref{discussion} the implications for our understanding of disk dispersal and the nature of the warm molecular disk surface layer are discussed.

\section{Observations}
\label{observations}
Spectro-astrometric observations were obtained as part of a large CRIRES survey of infrared molecular emission from protoplanetary disks and young stellar objects
within the framework of the European Southern Observatory (ESO) Large Program 179.C-0151 \citep{Pontoppidan11}. 
Spectro-astrometry is a highly sensitive method that allows a single telescope to obtain both spatial and kinematic information on gas-phase lines on very small 
scales, $<1$\,milliarcsecond, and at very high spectral resolution, $\lambda/\Delta \lambda \sim 100\,000$. The final accuracy of a spectro-astrometric measurement depends linearly
on both the signal-to-noise and the width of the spatial PSF. Basically, the method measures the spatial centroid offset of the spectrum
as a function of wavelength across a line or other spectral feature,
relative to the continuum. This approach can reveal spatial structure on scales much smaller than the
formal diffraction limit of the observation. Spectro-astrometry was first 
developed for chromatic imaging and spectroscopy for the detection of stellar binaries in the visible range using specialized instrumentation \citep{Beckers82, Christy83, Aime88}.
The modern form, using an echelle spectrograph, was first presented by \cite{Bailey98}, and is reviewed by \cite{Whelan08}. Infrared ($\lambda\gtrsim 1\,\mu$m) spectro-astrometry of molecular lines with CRIRES was introduced 
by \cite{Pontoppidan08}, who presented CO data from three transitional disks, TW~Hya, HD~135344B and SR~21. They showed that sub-milliarcsecond precisions could
routinely be achieved and that the basic geometries of the line emitting regions -- 
sizes, inclinations and position angles -- could be determined with a high degree of confidence. 
Here we extend this sample to a much wider range of disks in terms of stellar type and evolutionary stage.

\subsection{Observing strategy}
Targets were selected for spectro-astrometric observations according to overall brightness and line-to-continuum contrast, as well as to cover
as wide a range as possible in known disk and stellar characteristics.  The line-to-continuum contrast is a particularly important parameter to consider, since
the accuracy of the measured astrometric signal depends roughly linearly on this parameter (see \S\ref{Formulation}). 
However, we were successful in obtaining high-quality spectro-astrometry for
sources with line-to-continuum ratios spanning $(F-C)/C \sim 0.1-2$, where $F$ is the total flux and $C$ is the continuum flux. The central stars include spectral types
from K7 to late A, and cover luminosities of 0.1 to 100\,$L_{\odot}$. The targets also span a wide range in accretion rates from $\log \dot{M}\sim -9$ to $-6$\,$M_{\odot}\,\rm yr^{-1}$. 

Overall, the sample is intended to represent the diversity found among protoplanetary disks to the extent that is possible given the limitations on the size of the sample. 
A log of the observations is given in Table \ref{obs_table}, including those already described in \cite{Pontoppidan08}, while Table \ref{source_prop} summarizes 
the properties of the central stars. 

The lack of a correction for differential refraction between the effective wavelengh of the slit viewing camera, which usually operates at $H$- or $K$-band, and the spectral wavelength precluded the observation of low-elevation targets for much of duration of the campaign\footnote{A correction for differential refraction was
implemented in the observing software on November 28, 2008.}, including
sources in the Taurus and Chamaeleon star forming regions. As a consequence, most targets are located in Ophiuchus, Serpens, Lupus and Corona Australis. 

\begin{figure}
\includegraphics[width=8cm]{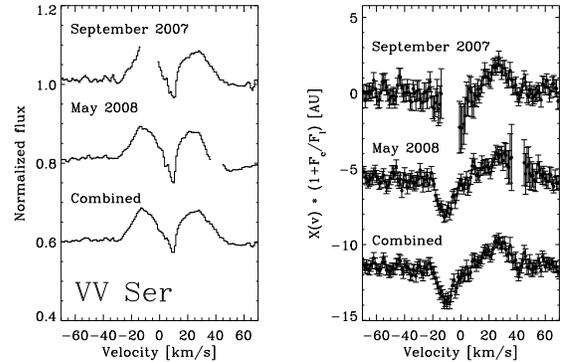}
\caption[]{Example of how observations at several epochs are combined to fill gaps left by saturated telluric lines. The flux spectra (left) and spectro-astrometry (right)
shown are of VV Ser taken at a slit $\rm P.A.=15\degr$.  }
\label{epoch_comb}
\end{figure}

For any given single observation, the wavelength coverage will not be complete due to the presence of saturated telluric absorption lines. This is 
particularly true for CO and water, for which the Earth's atmosphere absorbs in the same transitions as those targeted.
For this reason, a number of observations were repeated with a cadence of 3 months to more than a year. The primary purpose was to take advantage of the Earth's velocity around the Sun
to shift the telluric CO lines relative to those of the target disks. By observing targets at different epochs, complete line profiles could be constructed by combining of spectra obtained at each side of a given object's transit date. An example of how observations at two different epochs were combined to complete the spectral coverage of the CO lines is shown in Figure \ref{epoch_comb}. This strategy also allows for a shallow search for variability in the astrometric spectra, although the shifting telluric absorption often make sensitive comparisons difficult. Variability at such time scales may be expected since the Keplerian time scale at the radii traced, $\sim 0.1-1\,$AU is similar to or shorter than the cadence of observations. We can note that we did not detect any obvious variability in the sources observed during several epochs, but that we do not rule variability below the $\sim$20\% level. 

\begin{figure}
\includegraphics[width=8cm]{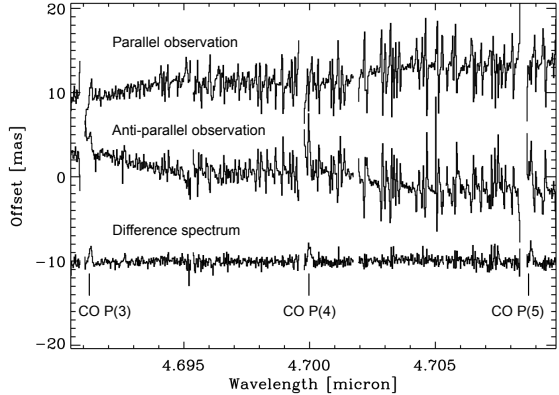}
\caption[]{Example of how parallel and anti-parallel slit positions are used to calibrate the astrometric spectra. Shown is the P.A.$=55\degr$ CO observation of AS 205N.
It is seen that a single spectrum is filled with astrometric artifacts, in this case mostly due to telluric O$_3$ and CO. Also apparent is a residual low frequency artifact due to
uncertainties in the distortion correction, which is also removed by the self-calibration, producing a flat astrometric final spectrum with an RMS of $\sim$0.5\,mas.}
\label{sa_calib}
\end{figure}

Significant artifacts in the astrometric spectra due to flat field and point spread function (PSF) effects may remain. The latter occurs if the PSF is not
exactly rotation symmetric - and PSFs never are. Indeed, \cite{Brannigan06} found
such artifacts to be a common feature of spectro-astrometric observations. The correction for PSF artifacts is therefore an
essential calibration of spectro-astrometry, without which meaningful analysis is not possible. To 
effectively correct for PSF artifacts, all astrometric spectra were obtained with the slit oriented at the desired position angle (P.A.), as well as at an antisymmetric $P.A.+180\degr$. In order to 
ensure that the instrument was kept as stable as possible, a special CRIRES observing template was developed in which the
grating and prism angle piezos were kept unchanged while the derotator changed the slit position angle. Furthermore, 
identical jitter patterns were used for the parallel and anti-parallel slit positions to maximize the reproducibility of
artifacts. The difference average between these two antisymmetric spectra  cancel out PSF artifacts while preserving any real signal.
The efficacy of the procedure is demonstrated in Figure \ref{sa_calib}.

\subsection{Data reduction}
\label{reduction}
The data were reduced using our own IDL scripts. The procedure includes flat-fielding, correction for the
non-linearity of the CRIRES detector response following the description in the CRIRES documentation issue 86.2, co-adding individual nod pairs and correction for spatial distortion. The flux
spectra were extracted using optimal extraction \citep{Horne86}, and were corrected for telluric absorption by division with a 
spectrum of an early-type standard star. The telluric standards were corrected for small airmass differences using a simple Beer law by minimizing
the telluric noise in a region of the spectrum relatively clear of intrinsic lines. The CRIRES grating position is not reproducible, so small relative shifts of order a few pixels in the dispersion direction
between the science and telluric spectra were applied. Finally, in some cases small differences in effective resolving power between target and standard star observation were corrected by degrading the 
resolving power by $0.1-0.2\,\rm km\,s^{-1}$ in either the science or telluric spectrum. Such differences may arise if differences in AO correction cause differences in the degree to which the source
fills the slit. No attempt was made to include an absolute flux calibration and all spectra were normalized to the continuum flux level. 

To further improve the signal-to-noise ratio, lines from similar transitions were averaged, wherever possible. 
In particular, this is important for the primary survey of the CO rovibrational band around 4.7\,$\mu$m, where typically 8 nearly identical (in terms
of energy and collisional rates) lines can be averaged in a single CRIRES setting. The uncertainty in the spectro-astrometric signals for typical sources is background-limited. 

\subsection{Spectro-astrometric formulation}
\label{Formulation}
For the spectro-astrometry, we find that the most numerically stable and reproducible centroid ($X$) estimator as a function of line velocity $v$ is of the form:
\begin{equation}
X(v) = K \frac{\sum_i (x_i(v)-x_0) F_i(v)}{\sum_i F_i(v)}, \mbox{ [pixels] }
\label{sadef}
\end{equation}
where $x_i$ is the spatial location of a pixel and $F_i$ is the flux contained in that pixel. 
The centroid $X(v)$ must also be corrected, using a constant factor $K$, for the fact that not all light will be included within the aperture. 
That is, $X(v)$ depends on the range over which $i$ is defined in the centroid estimator. In practice, however, $K$ is small, $< 1.5$, and is 
estimated using a modeled PSF measured on the continuum of the spectrum. Consequently,
the relative accuracy on the amplitude of the spectro-astrometric signal is likely $\lesssim 10\%$, which is confirmed by repeated
observations of the same sources (see also \S\ref{observations}). 

The uncertainty on the centroid is:

\begin{equation}
\sigma^2 (X) = K^2 \frac{\sum_j([j\sum_iF_i-\sum_i(iF_i)]^2 \times \sigma(F_j)^2)}{(\sum_iF_i)^4},
\label{sadef_unc}
\end{equation}
where the dependency of $X$ and $F$ on $v$ is left out for clarity. 

At this point, there is one more issue to consider. In our formulation, the flux is a sum of a continuum term and a line term: 

\begin{equation}
F_i=F_{C,i}+F_{L,i}.
\label{terms}
\end{equation}
Normally, one would be interested in determining the centroid offsets for the line term $F_L(v)$ only; the presence of the continuum term 
causes $X(v)$ to underestimate the true spatial offset. We call this effect {\it continuum dilution} \citep[see also][]{Pontoppidan08}. 
Algebraic manipulation of equations \ref{sadef} and \ref{terms} shows that the relation between the measured and true centroid offset is:

\begin{equation}
X_{\rm true}(v)=X_{\rm obs}(v)\times (1+F_C(v)/F_L(v)).
\label{cont_dilution}
\end{equation}
It is seen that the centroid naturally diverges for $F_L\rightarrow 0$. This makes it inconvenient to display the observed spectra in terms of $X_{\rm true}(v)$, and we
consequently display all spectra in terms of $X_{\rm obs}(v)$. The reader should thus be aware that the true spatial extent of the line emission is higher by a factor $1+F_C(v)/F_L(v)$.
In the following, we define a scalar {\it amplitude} of a given astrometric spectrum as the maximal value of the centroid offset: $A={\rm Max}(|X(v)|)$. 
This is a convenient model-independent observable that provides a measure of the size of the line emitting region.

\begin{deluxetable}{lllll}
\tablecaption{Log of observations}
\tablehead{
\colhead{Star}  & \colhead{PA} & \colhead{Obs. Date} & \colhead{Int. Time} & \colhead{Spectral Range}\\
\colhead{ }        & \colhead{ }    & \colhead{ }                & \colhead{minutes}  & \colhead{$\mu$m} 
}
\startdata
LkHa 330   & 0\degr & 29/12/2008  & 24& 4.660-4.770  \\ 
LkHa 330   & 60\degr & 29/12/2008  & 24& 4.660-4.770  \\
LkHa 330   & 120\degr & 29/12/2008  & 24& 4.660-4.770  \\
CW Tau      & 0\degr & 1/1/2009  & 24& 4.660-4.770  \\ 
CW Tau      & 60\degr & 1/1/2009  & 24& 4.660-4.770  \\
CW Tau      & 120\degr & 1/1/2009  & 24& 4.660-4.770  \\
DR Tau      & 0\degr & 14/10/2007  & 24& 4.805-4.901 \\ 
DR Tau      & 60\degr & 14/10/2007  & 24& 4.805-4.901  \\
DR Tau      & 120\degr & 14/10/2007  & 24& 4.805-4.901   \\
TW Hya      & 63\degr & 26/4/2007& 40& 4.660-4.770  \\
TW Hya      & 153\degr & 26/4/2007& 40& 4.660-4.770 \\
HD 135344B & 0\degr & 22/4/2007& 20& 4.645-4.755  \\
HD 135344B & 60\degr & 4/9/2007  & 20& 4.660-4.770  \\
HD 135344B & 120\degr & 5/9/2007  & 20& 4.660-4.770  \\
GQ Lup      & 0\degr & 2/5/2008  & 24& 4.660-4.770  \\
GQ Lup      & 60\degr & 2/5/2008  & 24& 4.660-4.770  \\
GQ Lup      & 120\degr & 2/5/2008  & 24& 4.660-4.770  \\
GQ Lup      & 0\degr & 4/8/2008  & 24& 4.660-4.770  \\
GQ Lup      & 60\degr & 4/8/2008  & 24& 4.660-4.770  \\
GQ Lup      & 120\degr & 4/8/2008  & 24& 4.660-4.770  \\
HD 142527& 60\degr & 7/8/2008  & 12& 4.639-4.749  \\
HD 142527& 150\degr & 7/8/2008  & 12& 4.639-4.749  \\
RU Lup       & 0\degr & 26/4/2007& 20& 4.660-4.770  \\
RU Lup       & 0\degr & 27/4/2008& 24& 4.660-4.770  \\
RU Lup       & 60\degr & 27/4/2008& 24& 4.660-4.770  \\
RU Lup       & 120\degr & 27/4/2008& 24& 4.660-4.770  \\
HD 144432& 6\degr & 2/8/2008  & 16& 4.639-4.749  \\
HD 144432& 66\degr & 2/8/2008  & 16& 4.639-4.749  \\
HD 144432& 126\degr & 2/8/2008  & 16& 4.639-4.749  \\
AS 205N      & 55\degr & 29/8/2007& 16& 4.660-4.770  \\
AS 205N       & 115\degr & 29/8/2007& 16& 4.660-4.770  \\
AS 205N       & 55\degr & 29/8/2007& 16& 2.905-2.977  \\
AS 205N       & 115\degr & 29/8/2007& 16& 2.905-2.977  \\
AS 205N       & 175\degr & 1/9/2007  & 24& 4.660-4.770  \\
AS 205N       & 55\degr & 2/5/2008  & 16& 4.660-4.770  \\
AS 205N       & 115\degr & 2/5/2008  & 16& 4.660-4.770  \\
AS 205N       & 175\degr & 2/5/2008  & 16& 4.660-4.770  \\
DoAr24E S   & 30\degr & 3/9/2007  & 20& 4.660-4.770  \\
DoAr24E S   & 90\degr & 2/9/2007  & 20& 4.660-4.770  \\
DoAr24E S   & 150\degr & 2/9/2007  & 20& 4.660-4.770  \\
SR 21         & 10\degr & 30/8/2007& 32& 4.660-4.770  \\
SR 21         & 70\degr & 30/8/2007& 32& 4.660-4.770  \\
SR 21         & 130\degr & 31/8/2007& 32& 4.660-4.770  \\
RNO 90      & 0\degr & 25/4/2007& 16& 4.660-4.770  \\
RNO 90      & 60\degr & 26/4/2007& 16& 4.660-4.770  \\
RNO 90      & 120\degr & 26/4/2007& 16& 4.660-4.770  \\
VV Ser       & 15\degr & 5/9/2007  & 20& 4.660-4.770  \\	
VV Ser       & 75\degr & 5/9/2007  & 20& 4.660-4.770  \\
VV Ser       & 15\degr & 1/5/2008  & 32& 4.660-4.770  \\
VV Ser       & 75\degr & 1/5/2008  & 32& 4.660-4.770  \\
VV Ser       & 135\degr & 1/5/2008  & 32& 4.660-4.770  \\
S CrA N       & 30\degr & 4/9/2007  & 20& 4.660-4.770  \\
S CrA N       & 90\degr & 4/9/2007  & 20& 4.660-4.770  \\
S CrA N       & 150\degr & 3/9/2007  & 20& 4.660-4.770  \\
R CrA         & 0\degr & 1/9/2007  & 12& 4.660-4.770  \\
T CrA         & 0\degr & 26/4/2007& 20& 4.660-4.770  \\
T CrA         & 90\degr & 26/4/2007& 20& 4.660-4.770  \\
\enddata

\label{obs_table}
\end{deluxetable}

\begin{deluxetable*}{llllllllll}
\tablecaption{Disk and stellar properties}
\tablehead{
\colhead{Source} & Class\tablenotemark{a} & line profile\tablenotemark{b}&\colhead{distance\tablenotemark{c}} & \colhead{$v_{\rm CO (LSR)}$}   & \colhead{$L_*$}        & \colhead{Sp. T.} & \colhead{$M_*$\tablenotemark{d}} & \colhead{$\dot{M}$\tablenotemark{e}}  & \colhead{references\tablenotemark{f}}\\
                           &                                       &                                               &\colhead{pc}                                      & \colhead{$\rm km\,s^{-1}$} & \colhead{$L_{\odot}$} &                          &  \colhead{$M_{\odot}$}                      &    \colhead{$M_{\odot}\,\rm yr^{-1}$}       &
}
\startdata

LkHa 330     & trans. disk & Keplerian      & 250 & 9.0 & 16   & G3      & 2.5  & -8.80/-8.80 & 2,12 \\
CW Tau        & CTTS          & self-abs.      &140  & 7.5 &0.8   & K3      & 1.2   & -8.80/-7.99 & 15,19\\  
DR Tau        & CTTS           & single-peak &140  & 10.7 &0.9 &  K5     &  1.0  & -7.5/-5.1& 16,17,18,19\\  
TW Hya        & trans. disk & Keplerian      & 51   & 3.0 &0.23 & K7      & 0.7  & -8.80/-8.80 & 14 \\ 
HD 135344B& trans. disk & Keplerian      &  84  & 7.5 &8     & F3       & 1.6  & -8.30/-8.30 &6\\
GQ Lup        & CTTS           & Keplerian     & 150 & 1.0 &0.8  & K7      &  0.8  & -8.00/-8.00 &11\\ 
HD 142527  & HAeBe        & Keplerian      &198  & 5.0 &69   & F6       &  3.5  & -7.16/-7.16 & 6\\ 
RU Lup         & CTTS          & single-peak & 150 & 3.5 &0.4  & K7       &  0.7  & -7.70/-7.70 & 5\\
HD 144432  & HAeBe        & Keplerian     & 145 & 6.0 &10   & A9       &  1.7  & 7.07/-7.07 & 6\\ 
AS 205N      & CTTS          & single-peak &125  & 4.5 &7.1  & K5        &  1.1  & -6.14/-6.14 & 10\\ 
DoAr 24E S   & CTTS          & self-abs.      & 125 & 3.5 &1.3  & K7-M0 &  0.7  & -8.46/-8.46 &8\\ 
SR 21           & trans. disk & Keplerian      & 125 & 3.0  & 15  & G2.5     &  2.2  & $<-8.84$&2\\ 
RNO 90        & CTTS         & Keplerian      & 125 & -1.5&4.0   & G5        & 1.5   & -- & 3\\ 
VV Ser          & HAeBe       & Keplerian     & 415 & 7.0&125 & B1-A3  &  3.0   & -6.34/-6.34 & 9,4\\ 
S CrA N        & CTTS         & single-peak &130  & 2.4 & 2.3   &  K3       & 1.5    & --            &13\\ 
R CrA           & HAeBe       & self-abs.      & 130 & 5: &100    & B8-F5  & 3.5:   & -7.12/-7.12 & 1\\ 
T CrA           & CTTS         & self-abs.      & 130 & 7: &8   & F0-F5      &  1.6:   & $<-8.20$ & 7
\enddata
\tablenotetext{a}{Type of disk -- can be classical T Tauri star (CTTS), Herbig AeBe star or transition disk }
\tablenotetext{b}{Type of CO rovibrational line profile as discussed in \S\ref{this_paper}.}
\tablenotetext{c}{The distances are based on the current best estimates to the parent young clusters of the disks, many of which are determined by parallax measurements of known cluster members. \citep{Dzib10, Torres09, Loinard08}, \citep{Torres09}. The distance to Corona Australis is well determined using the orbit solution for the eclipsing binary TY CrA \citep{Casey98}. 
One exception is HD 135344B, which has an uncertain distance of 84-140\,pc \citep{Grady09}.}
\tablenotetext{d}{The mass of the central star is estimated based on the luminosity and spectral type using the evolutionary tracks of \cite{Siess00}.}
\tablenotetext{e}{Range of mass accretion rates found in the literature.}
\tablenotetext{f}{References used for the stellar properties and the mass accretion rates. }

\tablerefs{ [1] \cite{Bibo92}, [2] \cite{Brown07}, [3] \cite{Chen95}, [4] \cite{Dzib10}, [5] \cite{Herczeg08}, [6] \cite{Meijer08}, [7] \cite{Meyer09}, [8] \cite{Natta06},
[9] \cite{Pontoppidan07a}, [10] \cite{Prato03}, [11] \cite{Seperuelo08}, [12] \cite{Salyk09}, [13] \cite{Schegerer09}, [14] \cite{Thi10}, [15] \cite{White01}, [16] \cite{Mora01}, [17] \cite{Muzerolle03}, 
[18] \cite{Gullbring00}, [19] \cite{Johns-Krull02}}
\label{source_prop}
\end{deluxetable*}

\begin{figure*}
\centering
\includegraphics[width=17cm]{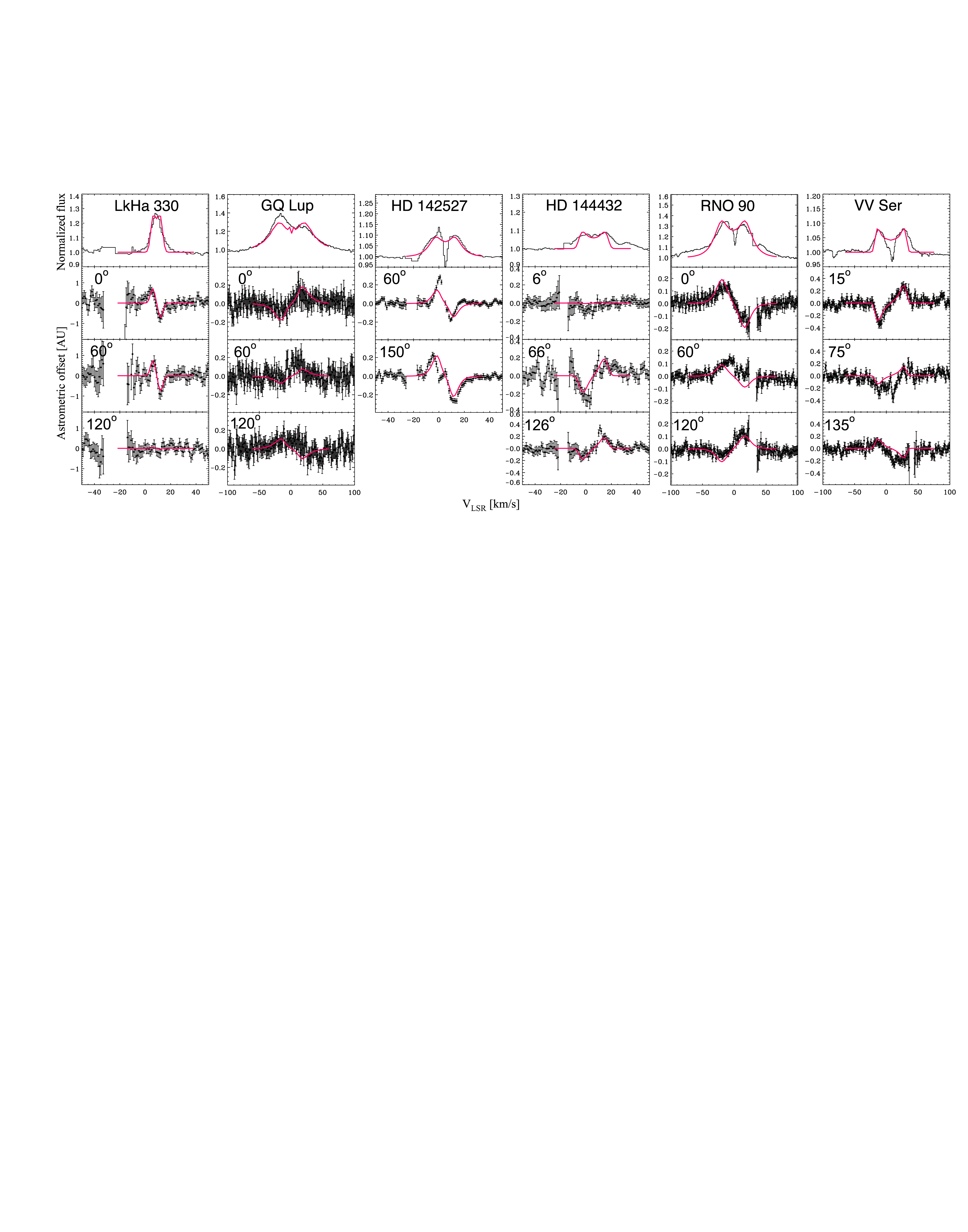}
\caption[]{Keplerian model fits using the simple flat-disk model superimposed on the data. The flux spectra are shown on top, while the lower panels
show the spectro-astrometry at different slit position angles. The red curves show the model fits. Note that the y-axes are given
in continuum-diluted units, and the real physical extent probed by the spectro-astrometry is larger by factors of ($1+F_C/F_L$). }
\label{kepler_models}
\end{figure*}
 
\begin{figure}
\includegraphics[width=4.2cm]{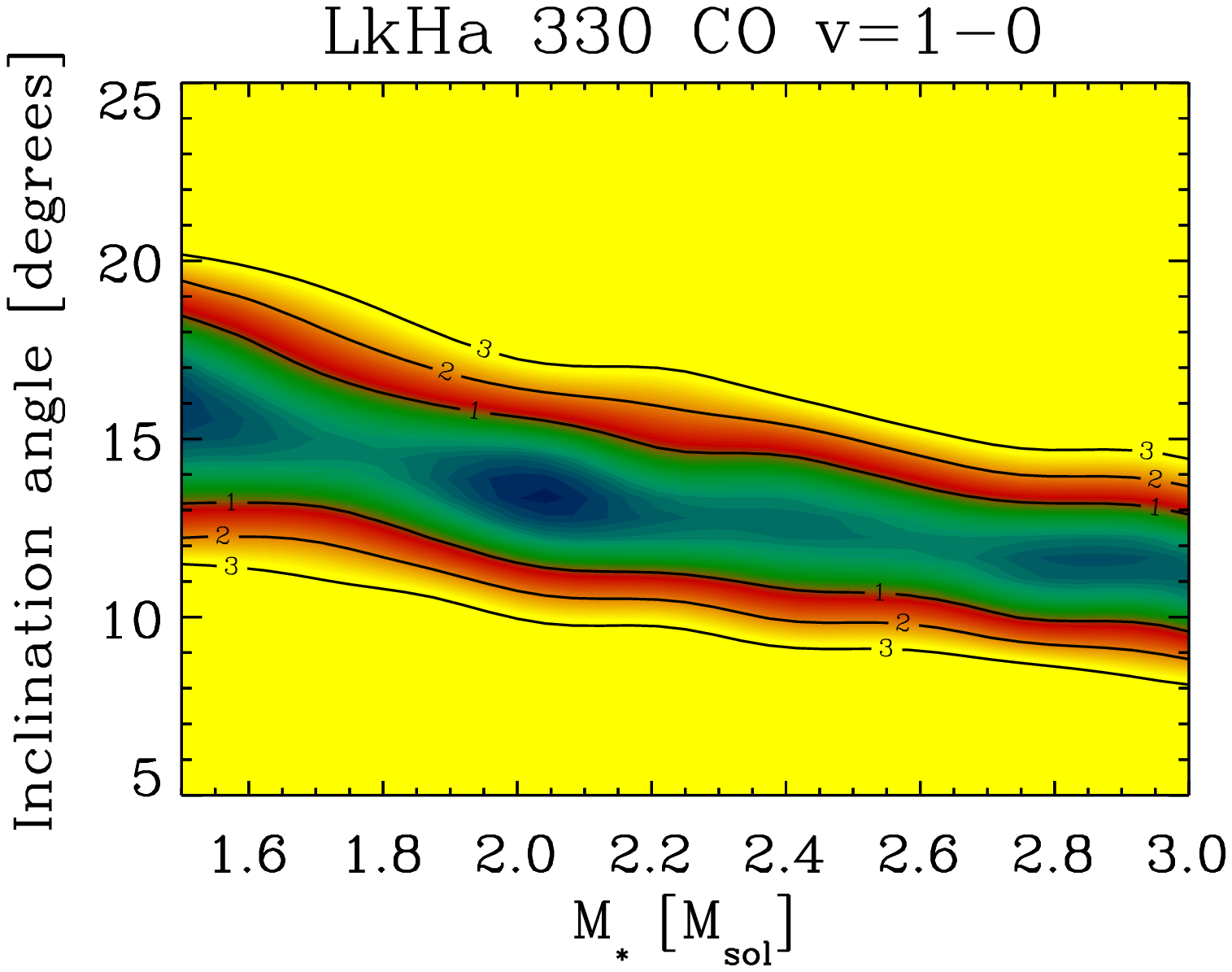}
\includegraphics[width=4.2cm]{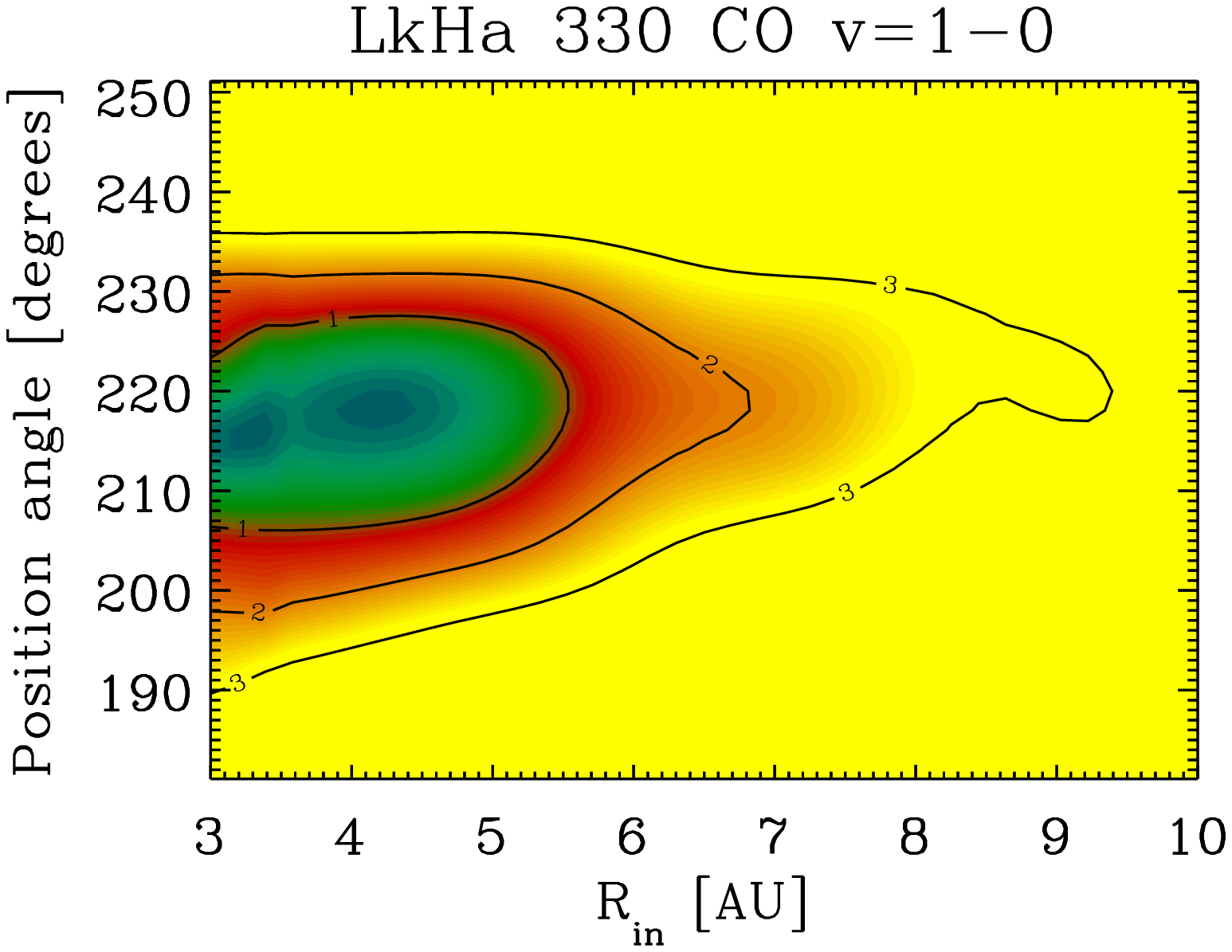}
\includegraphics[width=4.2cm]{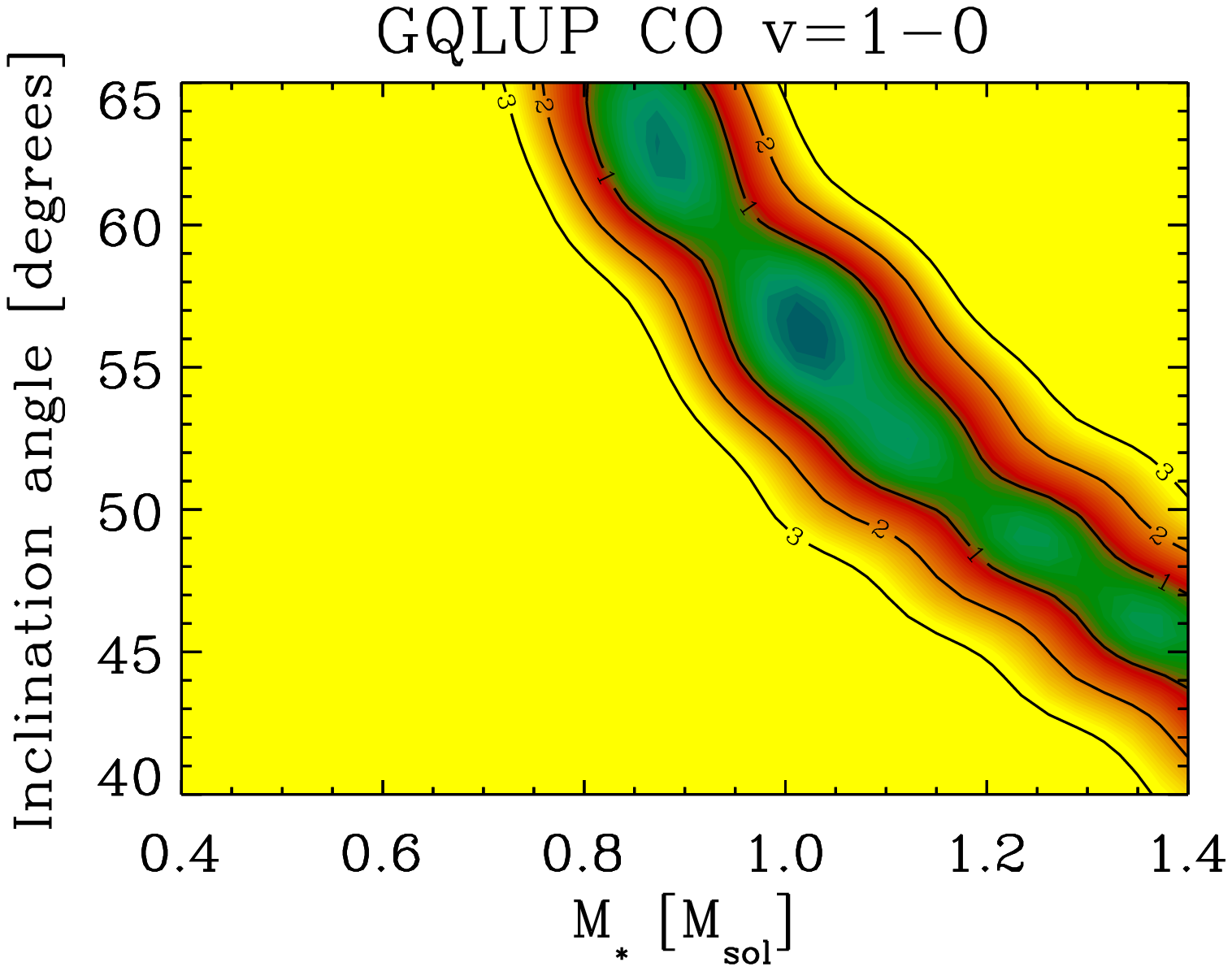}
\includegraphics[width=4.2cm]{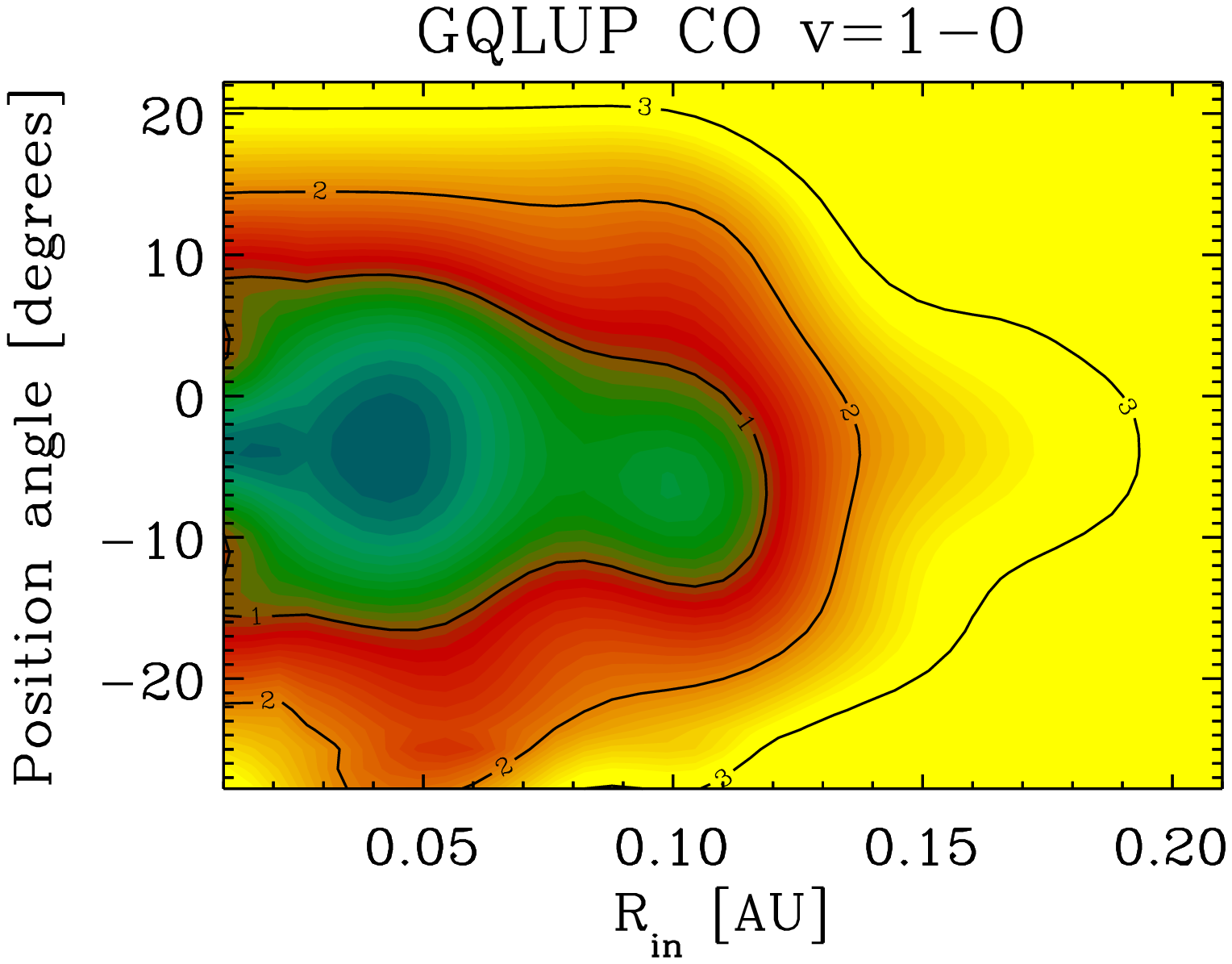}
\includegraphics[width=4.2cm]{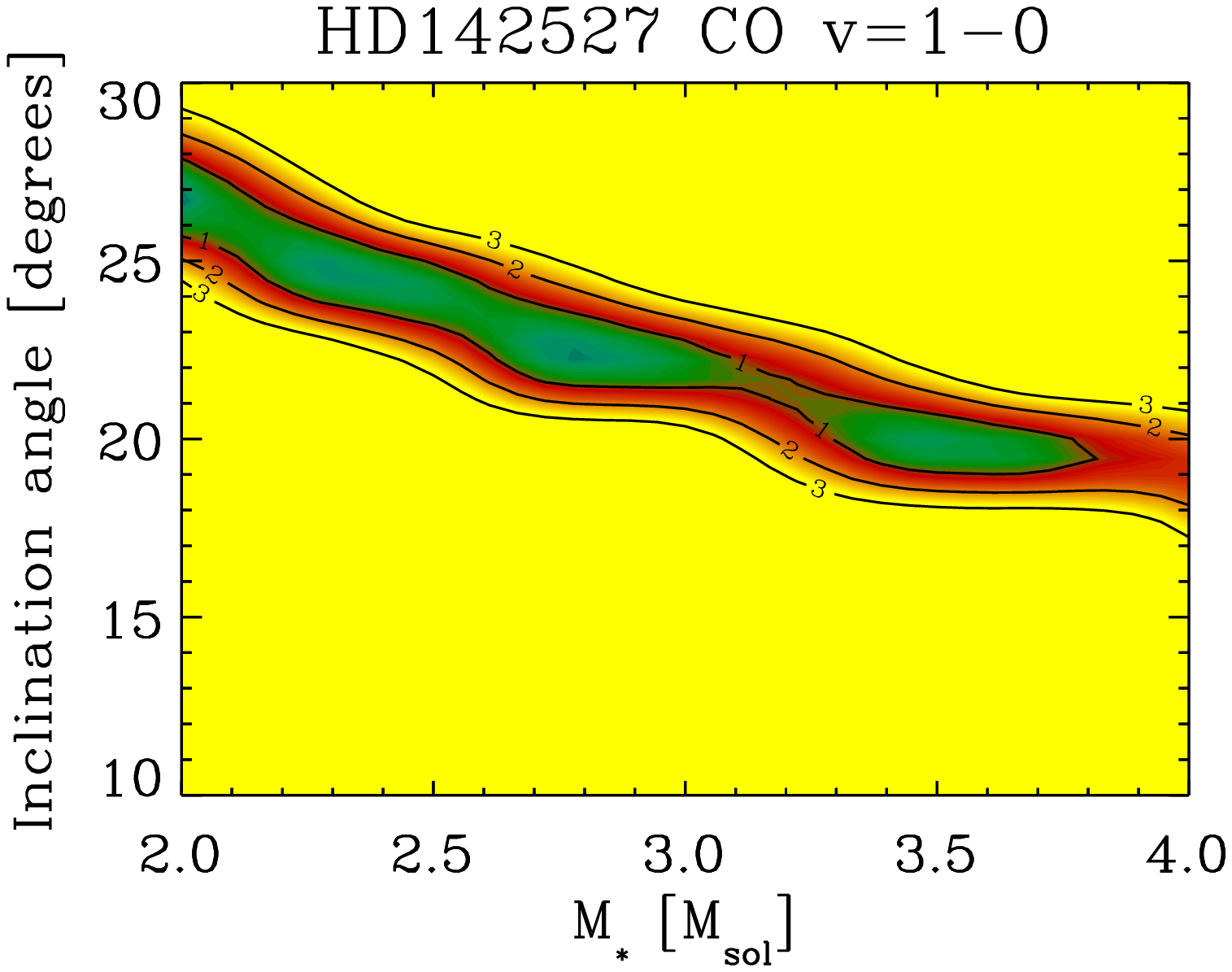}
\includegraphics[width=4.2cm]{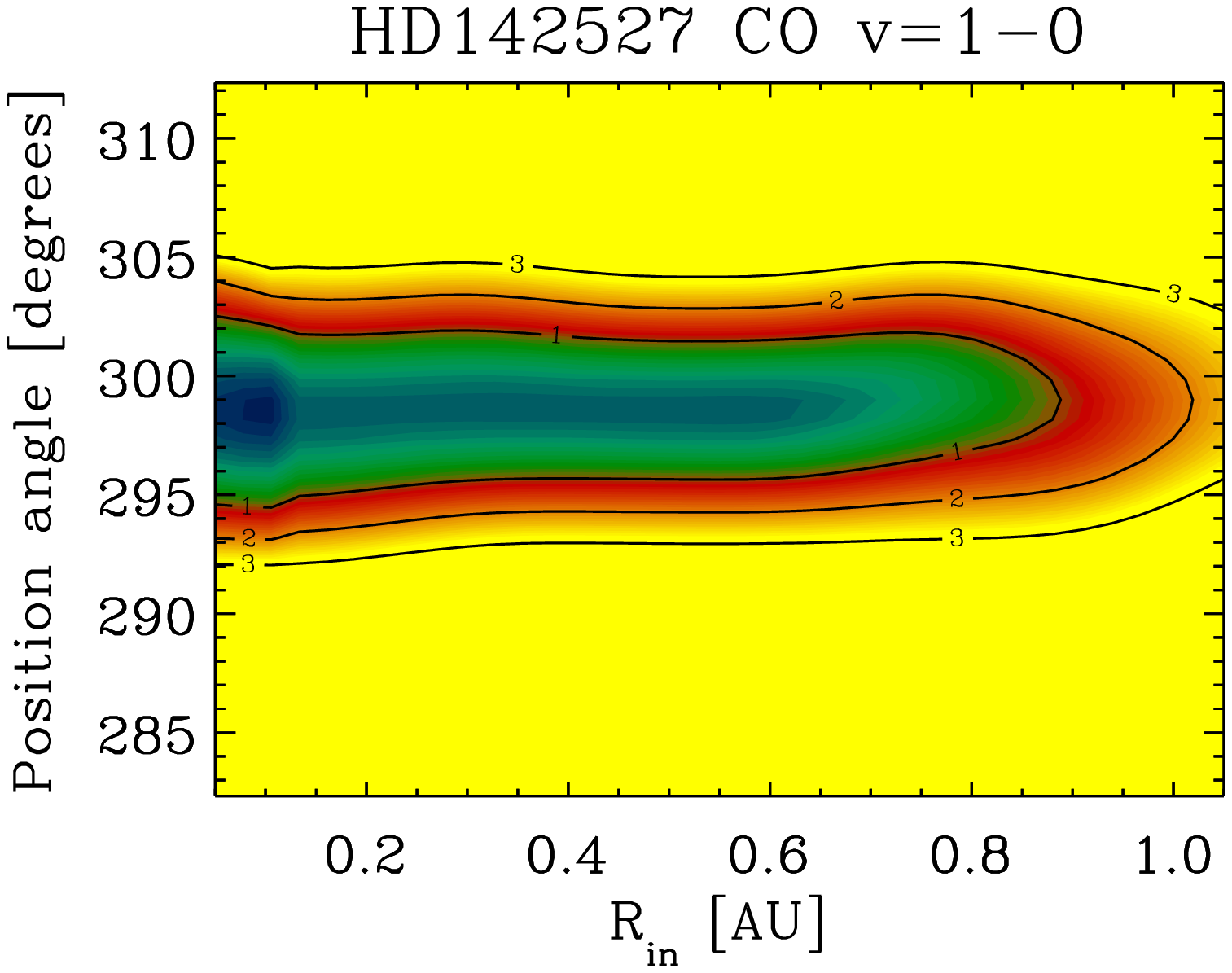}
\includegraphics[width=4.2cm]{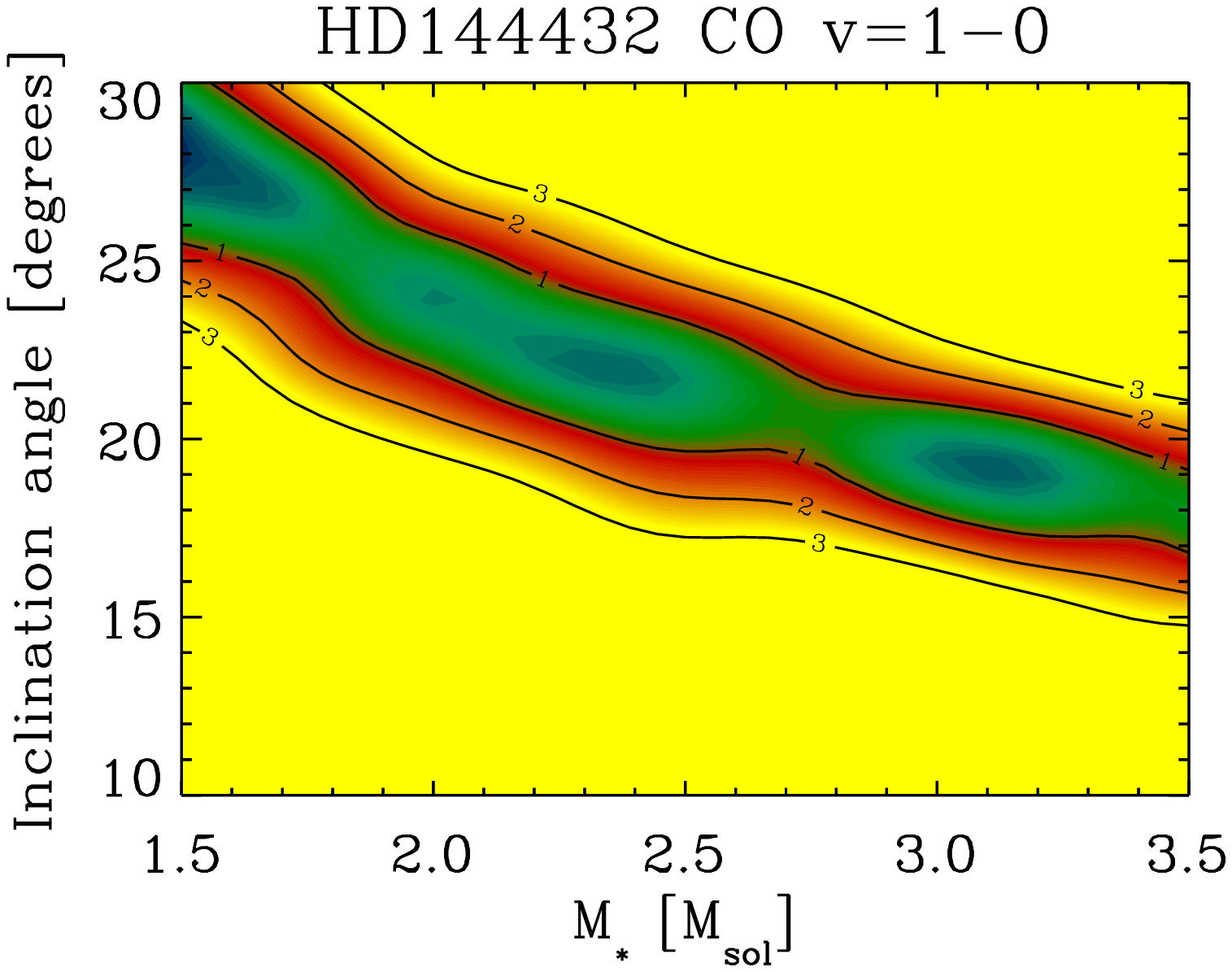}
\includegraphics[width=4.2cm]{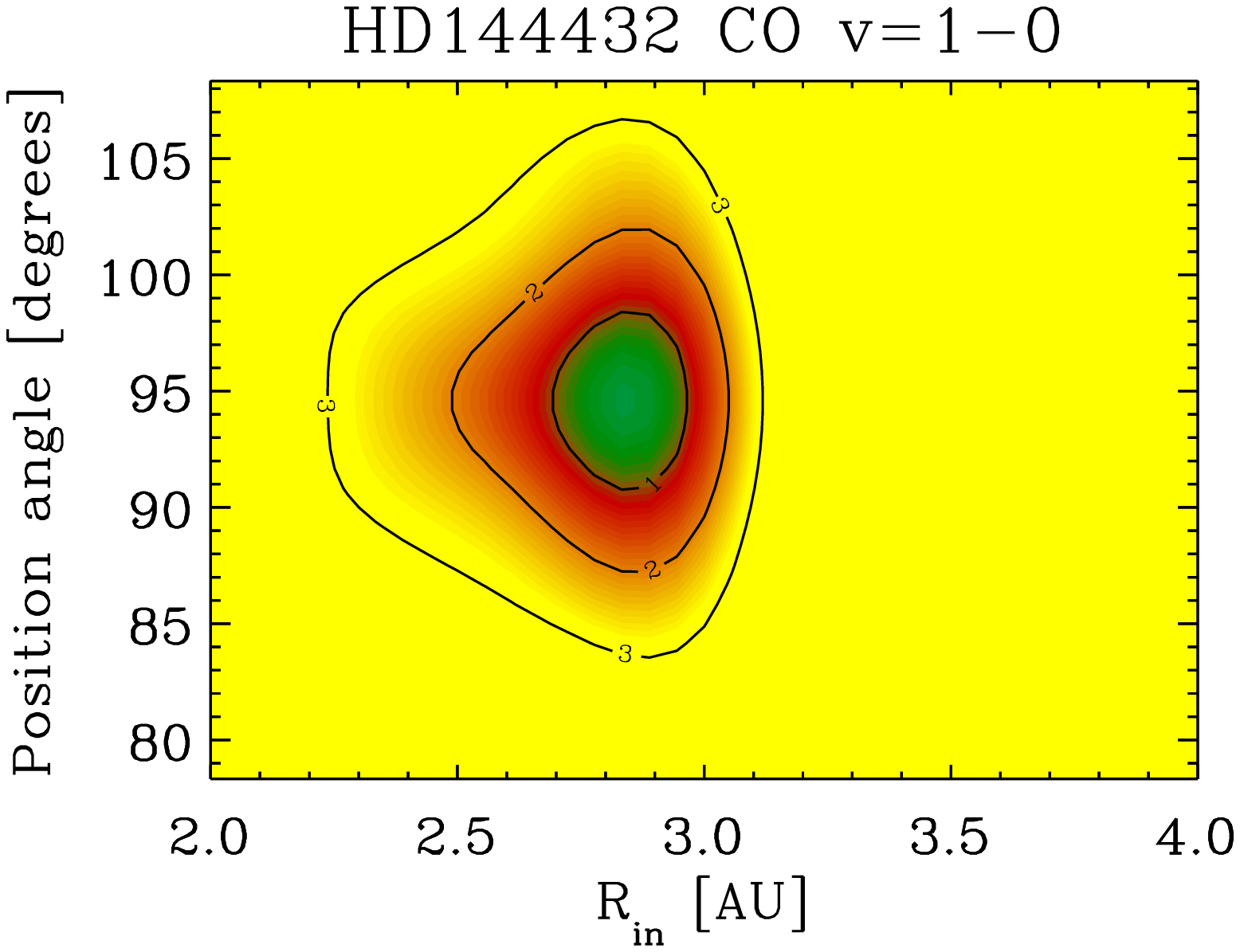}
\includegraphics[width=4.2cm]{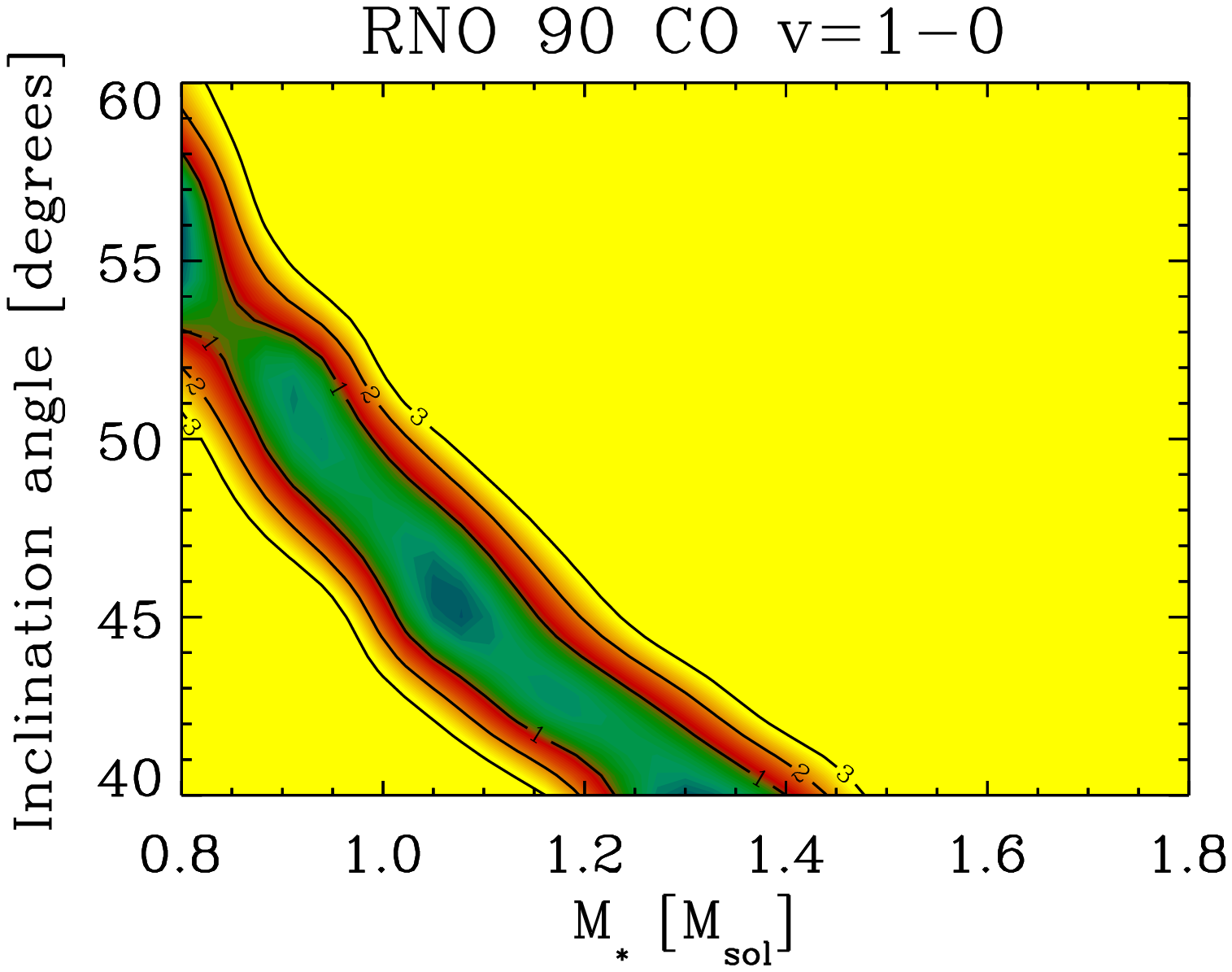}
\includegraphics[width=4.2cm]{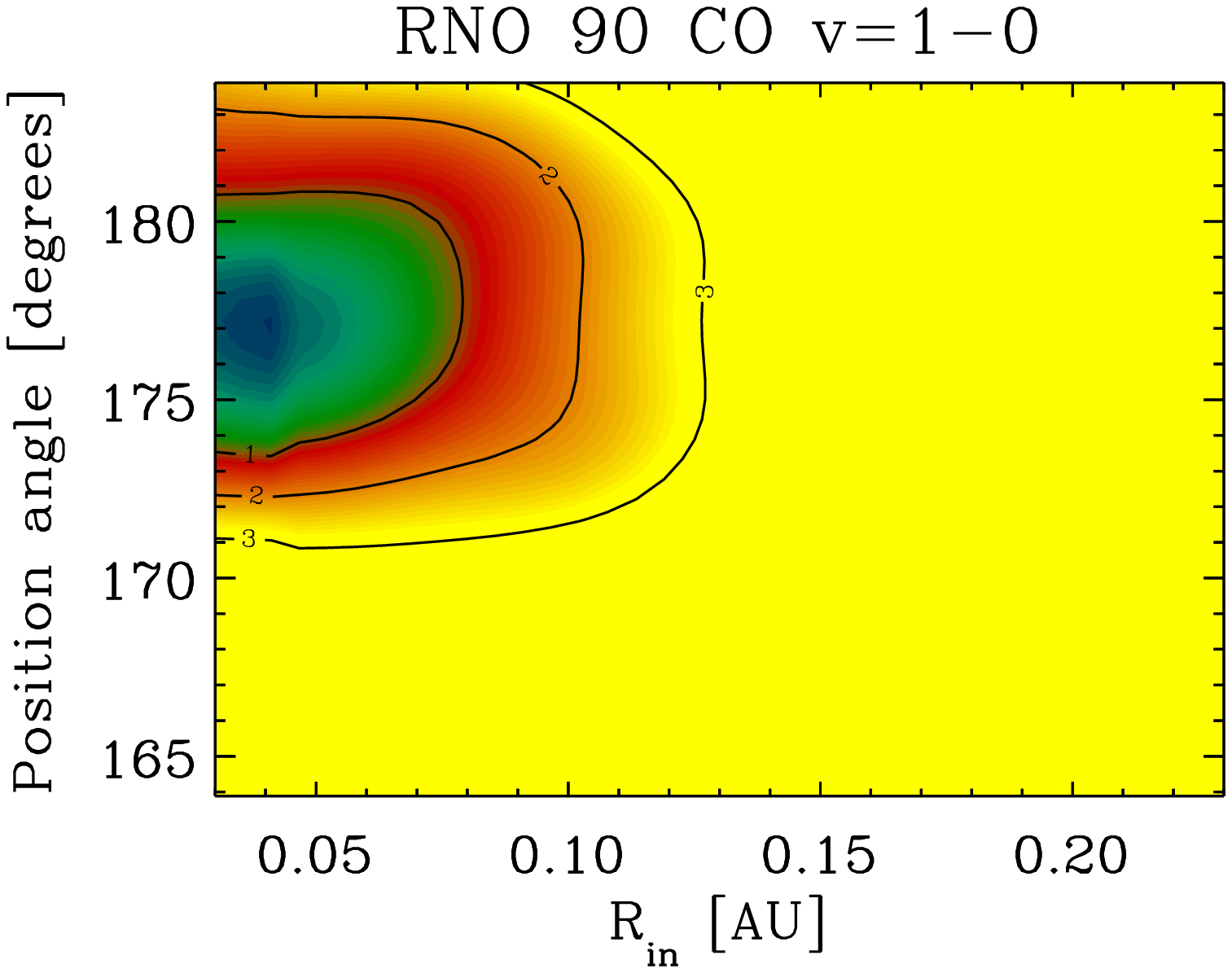}
\includegraphics[width=4.2cm]{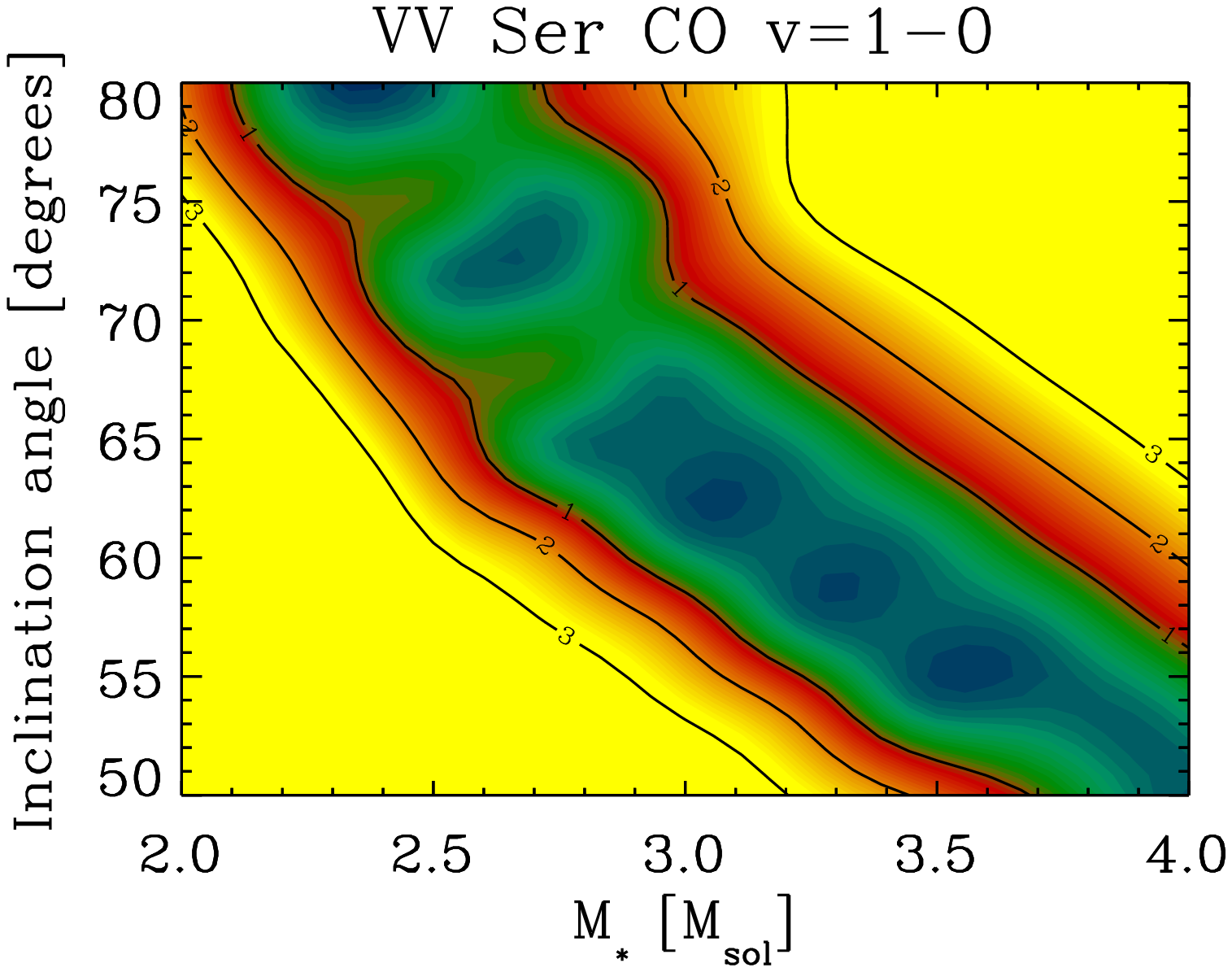}
\includegraphics[width=4.2cm]{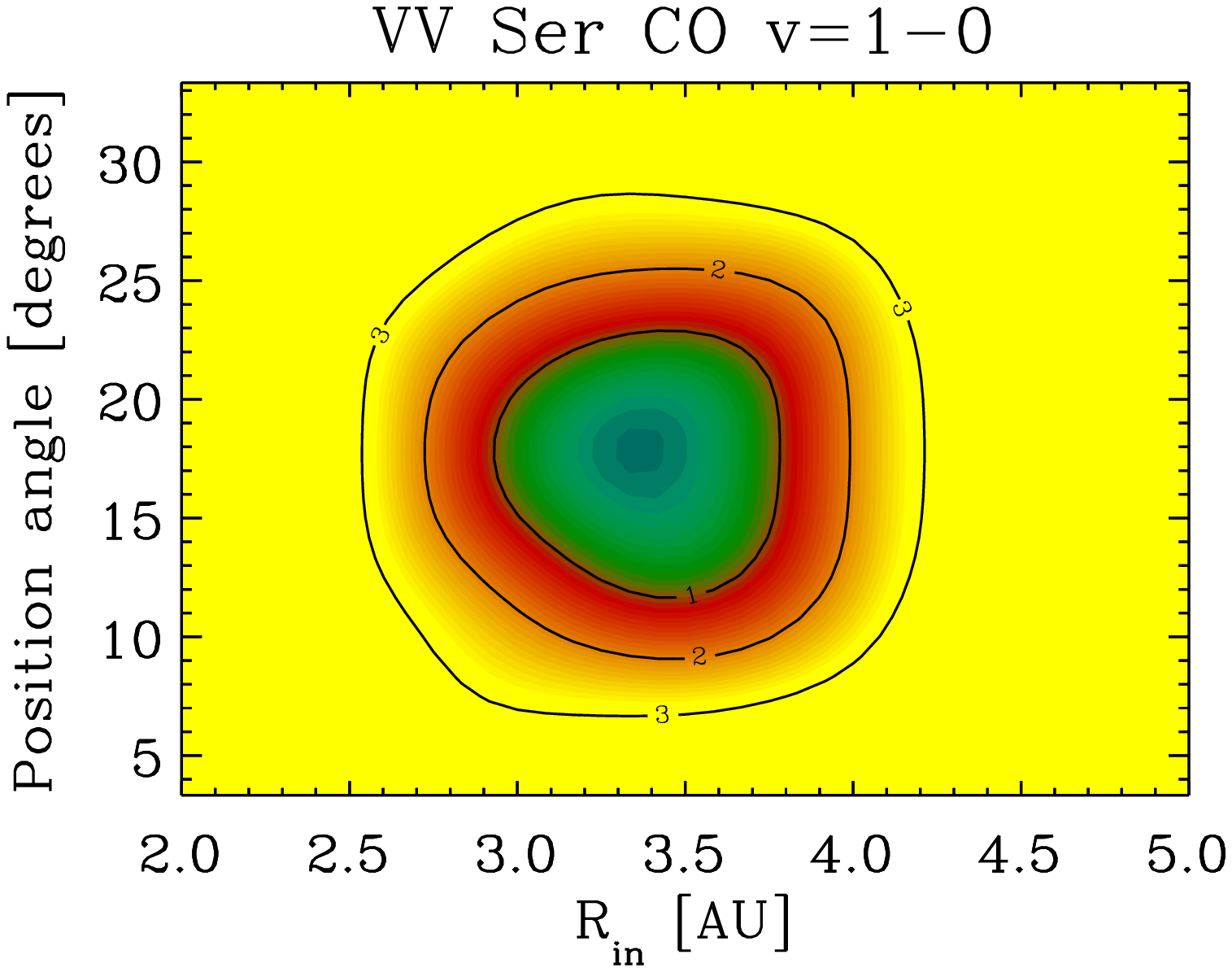}
\caption[]{Goodness-of-fit surfaces, using the $\chi^2$ statistic for the simple Keplerian model fits to the Keplerian sources \citep[see also][]{Pontoppidan08}. }
\label{kepler_chis}
\end{figure}

\section{Keplerian sources}
\label{Kepler_sources}

\subsection{Simple geometric models}
Keplerian sources are characterized by double-peaked line profiles in combination with broad astrometric spectra that display an anti-symmetric pattern
at all slit position angles. For disks viewed at inclinations close to face-on, the double-peak may blend into a single peak, but
if the astrometric spectra show an antisymmetric structure at all position angles, they are still considered Keplerian. 
The three disks discussed in \cite{Pontoppidan08}, HD 135344B, SR 21 and TW Hya, are in this category. Prototypical disks with clean double-peaked structure
include GQ Lup, RNO 90 and VV Ser. The astrometric and line flux spectra of the Keplerian disks are shown in Figure \ref{kepler_models}. 
As shown in \cite{Pontoppidan08} this structure is well explained by
a simple model of a radial, flat distribution of gas in circular orbits around a point mass, and the antisymmetric structure
is due to the relative spatial displacement of red- and blue-shifted gas in an inclined disk. If the slit is oriented along the major axis of the projected disk, 
the maximal astrometric displacement amplitude is seen. Conversely, for an exactly axisymmetric and flat disk, a slit
oriented along the disk minor axis produces no astrometric signal. Given at least two slit position angles, the disk position angle
can be determined with a high degree of confidence. The dominant line emitting radii can be determined, using the 
maximal astrometric offset for a slit aligned along the disk major axis. Finally, the disk inclination can be determined with confidence if a stellar mass is assumed and vice versa. 

We use the same simple geometric model as \cite{Pontoppidan08} to fit the data and determine the basic geometric parameters for the Keplerian disks. The parameters varied
are the inner emitting radius, the stellar mass, the disk inclination, $i$, and position angle, P.A. Figure \ref{kepler_chis} shows the
resulting goodness-of-fit contours for these four parameters. Table \ref{kepler_table} summarizes the best-fit parameters.
As expected, the stellar mass and the disk inclination are degenerate, but in such a way that even an uncertain assumption of the 
stellar mass allows an accurate determination of the disk inclination. The disk position angles are absolutely determined, while
the inner radius is somewhat dependent on the choice of the $T(R)$ relation (here assumed to be a power law with exponent $q=-0.5$). In lieu of the inner radius determined from the fit, 
the size of the line emitting ring, in an averaged sense, can also be estimated by using the amplitude $A$, defined above, and correcting for continuum dilution (see \S\ref{Formulation}). 

\subsection{A size-luminosity relation for ro-vibrational CO}
\begin{figure}
\centering
\includegraphics[width=8cm]{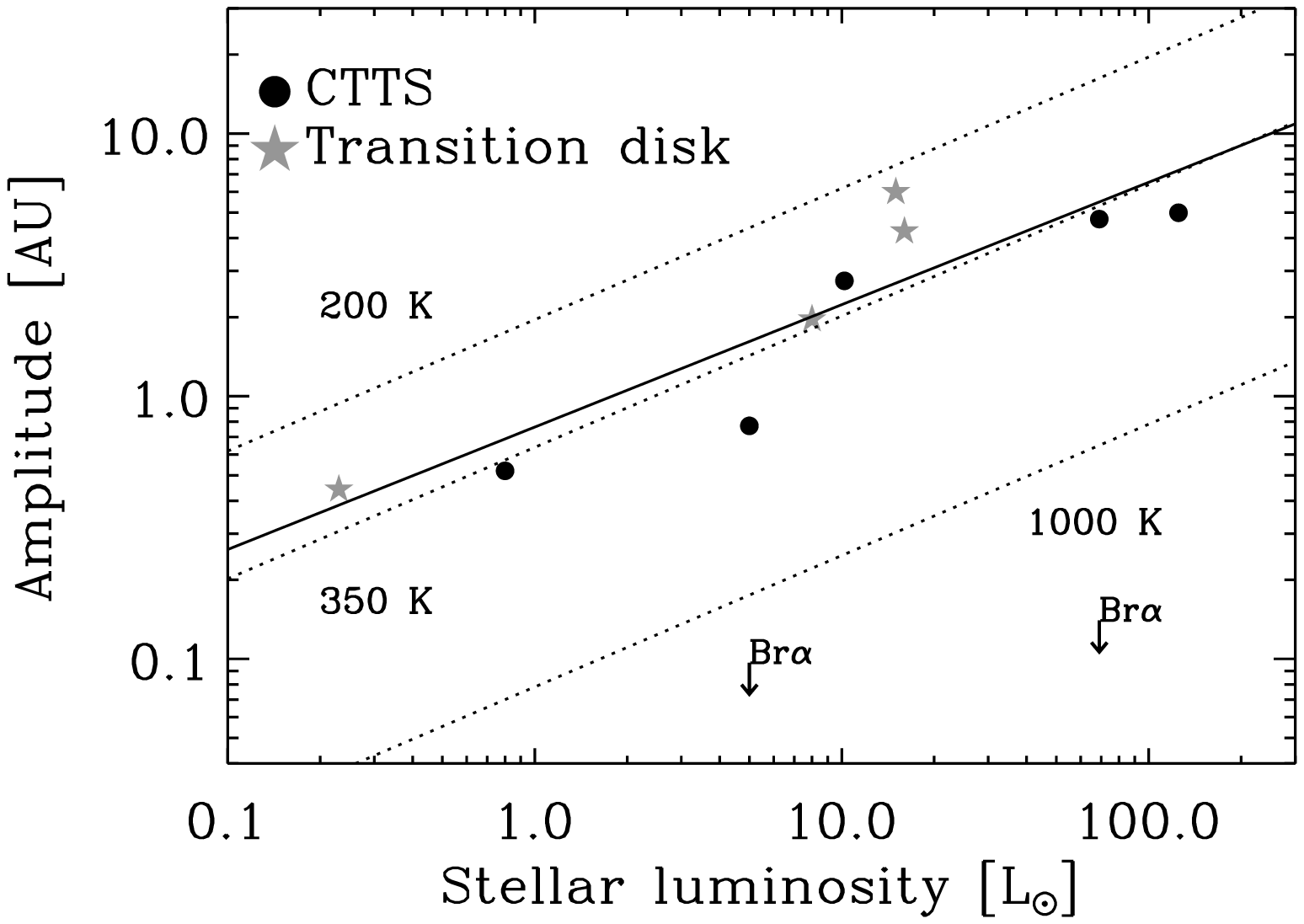}
\includegraphics[width=8cm]{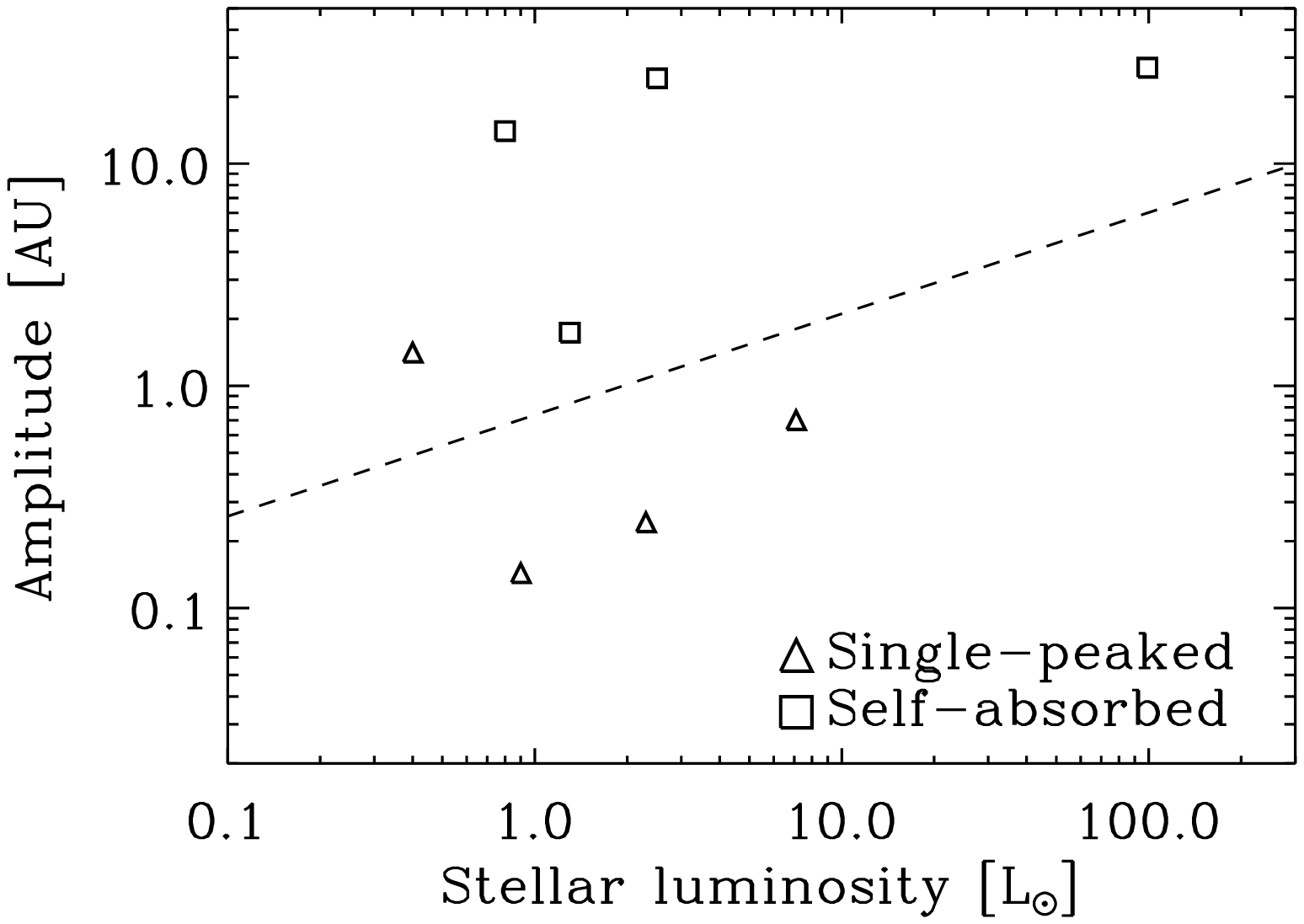}
\caption[]{Relation between the spectro-astrometric amplitude (continuum-corrected) and stellar luminosity for the four classes of astrometric spectra. The astrometric
amplitudes have been corrected for continuum dilution and are normalized to a distance of 125\,pc. Top: The Keplerian profiles divided
into the three transition disks from \cite{Pontoppidan08} and the CTTS's from this paper. The Keplerian disks exhibit a tight correlation with $A\propto L_{*}^{0.5}$. 
Bottom: The same as the top plot, but for the single-peaked and self-absorbed sources. No correlation is seen. Further, the self-absorbed sources show much greater amplitudes than
the Keplerian disks. }
\label{size_lum}
\end{figure}

In Figure \ref{size_lum}, the astrometric amplitudes are plotted versus the stellar luminosities. It is seen that the Keplerian disks show a 
clear correlation across nearly four decades in stellar luminosity as $A\propto L_*^{\alpha}$, with a best-fit exponent of $\alpha=0.48$. This is the relation expected for 
the radius of a specific equilibrium temperature as a function of stellar luminosity \citep[e.g.,][]{Dullemond01,Monnier02}. A very similar
size-luminosity relation was found for the $K$-band continuum emission from Herbig Ae/Be stars using the Keck interferometer \citep{Monnier05}.
Here, we find that the molecular gas obeys a similar relation, but on larger scales and extending down to sub-solar luminosities. 

The observed points are compared to the radii of different optically thin grey dust equilibrium temperatures, $R_{S}=1.1\sqrt{\epsilon_Q}(L_*/10^3~L_{\odot})^{1/2}(T_s/1500~K)^{-2}$ \citep{Monnier02}, where $\epsilon_Q$ is the ratio of the dust absorption efficiencies at the color temperatures of the incident and reemiiting radiation fields. Note that
the dust temperature also depends on additional radiation sources, such as that of the surrounding disk \citep{Dullemond01}, which will tend to push the radius at a specific
temperature outwards. However, this prescription allows a direct comparison to the analysis of \cite{Monnier05}.  

As is seen in Figure \ref{size_lum}, the gas lines are indeed dominated by gas at radii well beyond the interferometrically measured 
dust sublimation radii at $T$$\sim$$1000-1500\,$K, and consistently match dust at 350\,K. 
In the limit of a disk truncated at a sharp, optically thick, inner disk rim, the CO sizes correspond to dust at somewhat higher temperatures of $\sim$500\,K. 
It is important to realize that some molecular gas may still extend inwards, as indicated by the best-fit inner radii and the high velocities of emission
in the line wings \citep{Salyk07}, but the lines are not dominated by that component; the astrometric sizes measure the bulk of the gas emission.

Of particular interest, however, is that some transition disks fall on the size-luminosity relation defined by the classical disks, specifically TW Hya and HD 135344B -- 
SR 21 falls somewhat above the relation. This is consistent with
the findings of \cite{Salyk07}, \cite{Pontoppidan08} and \cite{Salyk09}; that the inner disk gas of some transition disks often has not been removed
in the same way as the population of small dust grains. For these disks, it now appears that there is not even a difference in the radii forming the CO gas emission, 
and that the lines simply follow the expected dependence on luminosity. This can be interpreted as the removal of the small dust opacity component
through the process of planetesimal formation, as this will preserve a significant column of gas in the inner evolved zone of the disk. The caveat is that this is a very small sample
of transition disks, and the inclusion of additional disks may show a greater degree of complexity, including the operation of other mechanisms of
inner disk clearing, such as photoevaporation or clearing by an unseen, but massive, (stellar) companion, both of which would tend to move the inner edge of the
gas-disk outwards. It will be interesting to see how many transition disks, in fact, fall on the relation. 

\begin{table}
\centering
\caption{Best fit model parameters for Keplerian sources}
\begin{tabular}{llll}
\hline
\hline
Star  & $R_{\rm in}$ [AU] & P.A. & $i$\tablenotemark{a} \\
\hline
LkhA 330    &  4$\pm$1         & 218$\pm$10$\degr$& 12$\pm 2 \degr$ \\       
GQ Lup        &   $<0.1$           & 357$\pm$10$\degr$& 65$\pm 10\degr$ \\
HD 142527 &  0.2$\pm$0.3   &299$\pm$3$\degr$ & 20$\pm 2\degr$ \\
HD 144432 &  2.7$\pm$0.1   &95$\pm$3$\degr$ & 25$\pm 3\degr$ \\
RNO 90       & 0.06$\pm$0.01 & 177$\pm$3$\degr$ & 37$\pm 4\degr$  \\
VV Ser        &  0.3$\pm$0.3     &17$\pm$4$\degr$  & 65$\pm 5\degr$ \\
\hline
\end{tabular}
\tablenotetext{a}{Assuming the stellar masses from Table \ref{source_prop}}

\label{kepler_table}
\end{table}

\subsection{Spectro-astrometry of hydrogen recombination lines in Keplerian sources}

Spectro-astrometry suffers from a fundamental symmetry ambiguity. Because the line centroid offsets are
measured relative to the continuum, departures from circular symmetry in the continuum brightness distribution
will be imprinted in the astrometric signal. For instance, if the continuum emission is due to a sharp inner rim in the
dust disk \citep[as in the models of][]{Dullemond01} and the disk is viewed at some inclination, an asymmetry should be 
seen in the astrometric line spectrum when the slit is oriented along the disk minor axis. However, it is fundamentally
not possible to disentangle this effect from an azimuthal asymmetry in the line intensity. In other words, CO spectro-astrometry
cannot distinguish between structure in the spatial distribution of line intensity and the spatial distribution of continuum intensity. 
This could be remedied if there were an independent line tracer of the stellar location. In this section, we suggest that hydrogen recombination
lines may, for many young stars, constitute such a tracer that may allow for spectro-astrometry of the continuum. 

There is a growing consensus that a dominant contributing process to H~I emission from T Tauri and Herbig Ae stars is 
magnetospheric accretion \citep{CalvetHartmann92, Muzerolle98, Bary08}, a shift from an original interpretation in the framework of stellar winds 
\citep{Hartmann90, Calvet92, Grinin91}, based on observed P Cygni profiles of the Balmer lines in some sources. While winds likely do contribute
to the optical H~I lines, near-IR lines tracing higher energies and densities appear to be dominated by accretion flows, as
indicated by a general lack of blue-shifted absorption \citep{Folha01}. If the lines are indeed dominated by accretion flows, 
the astrometry is expected to constrain the H~I emission to within a few stellar radii, corresponding to scales significantly smaller than
the disk co-rotation radii.

\begin{figure}
\centering
\includegraphics[width=7cm]{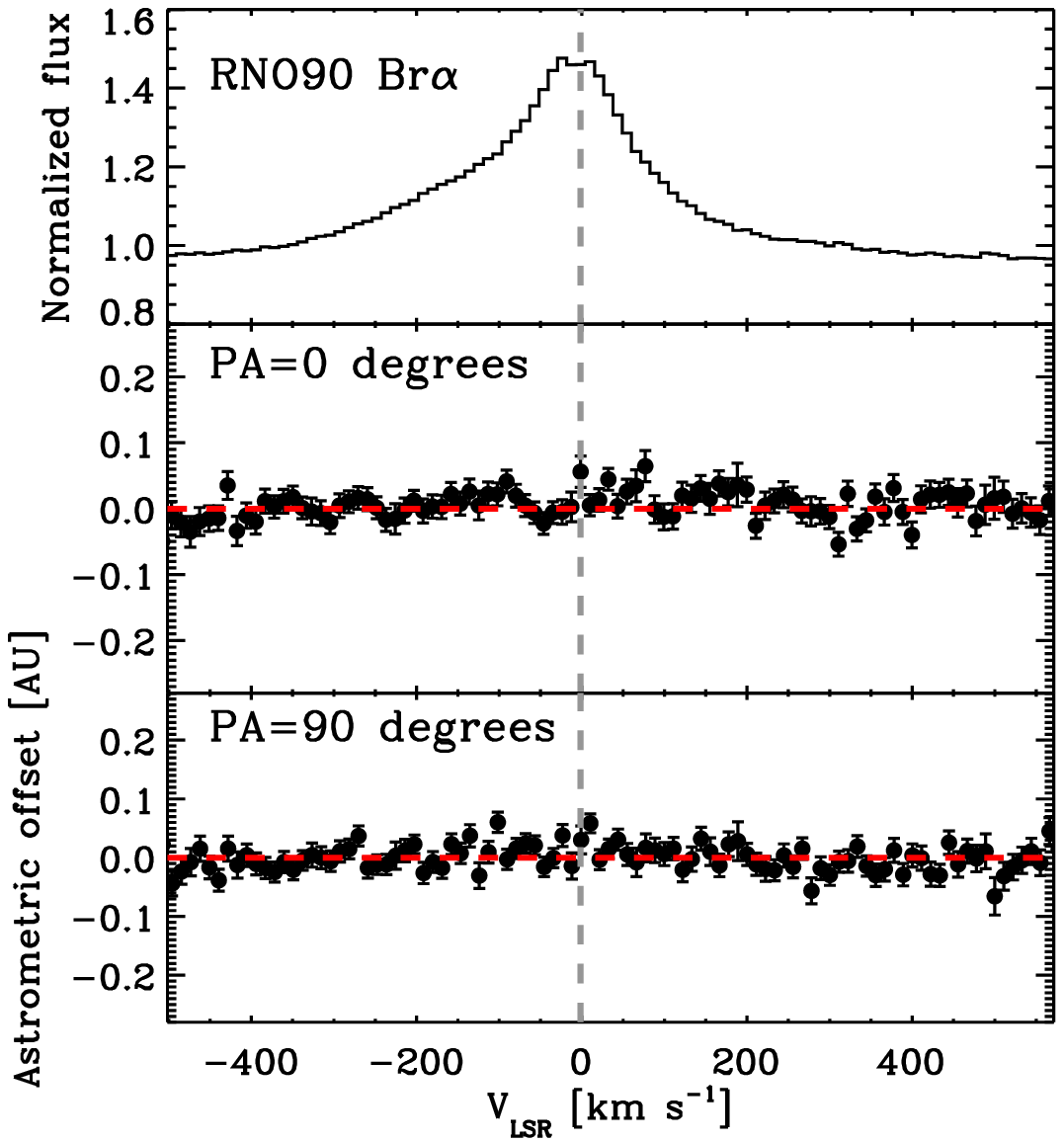}
\includegraphics[width=7cm]{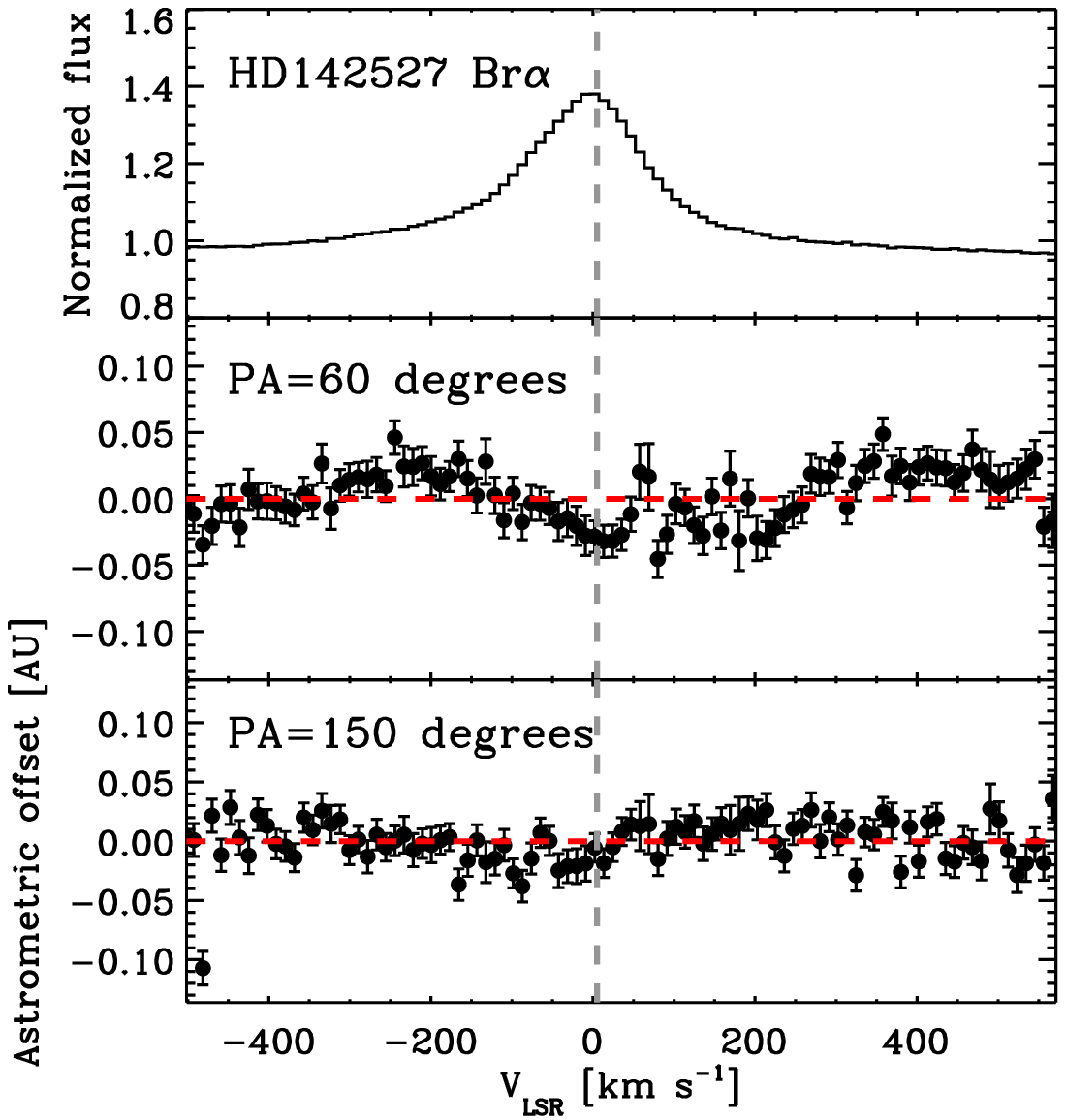}
\caption[]{Spectro-astrometry of the H~I Brackett $\alpha$ lines for RNO~90 and HD~142527. Because the H~I lines are highly over-resolved at
the CRIRES resolution, the spectra have been rebinned to a sampling of 12\,$\rm km\,s^{-1}$ to maximize the signal-to-noise. The formal
noise on the HD142527 astrometric spectrum is $\sim$0.13 mas, or $\sim 0.1\,$AU at a distance of 198\,pc, including continuum dilution. The vertical dashed lines indicate
the velocities of the CO rovibrational line centers.  }
\label{Bralpha}
\end{figure}

We obtained spectro-astrometry along two perpendicular position angles of the Brackett $\alpha$ lines at 4.05\,$\mu$m  for two of the Keplerian sources: RNO 90 and HD 142527. 
This line traces somewhat lower energies than e.g. the Brackett $\gamma$ line at shorter wavelengths, 2.16\,$\mu$m, but was chosen because the 
4\,$\mu$m continuum source is more likely to be comparable to that at 4.7\,$\mu$m. In particular, it is more likely to trace dust emission, which
is not necessarily the case at 2\,$\mu$m, where gas opacity may dominate \citep{Eisner09}. The resulting spectra
are shown in Figure \ref{Bralpha}. No astrometric signals are detected to limits of 0.35\,mas for RNO 90, while a tentative astrometric signal is seen at the 0.2-0.3\,mas level at a P.A$=60\degr$ for HD 142527. 
The formal displacement errors in the two sources correspond to similar physical sizes since HD 142527 is at almost twice the distance of RNO 90 (198\,pc versus 125\,pc). 
The conclusion is that the dust continuum emission is azimuthically symmetric and spatial asymmetries seen in the CO emission is not due to spatial structure
of the continuum emission. 

These results are consistent with an accretion origin of the infrared H~I lines. Alternatively, formation in a very compact stellar 
wind cannot be ruled out (a slight blue-shifted asymmetry is seen in the line profiles). Note that a potential stellar wind giving rise to H~I lines should not 
be confused with the much more extended disk wind discussed below in the context of the CO lines. More importantly, in the context of the inner disk, 
the lack of strong astrometric signatures from the H~I lines
rule out a sharp, inclined inner edge structure of the 4\,$\mu$m continuum emission. This is a result similar to that found with near-IR interferometry
of the 2\,$\mu$m continuum. It is interesting to note that \cite{Whelan04} found highly extended (10-100\,mas) spectro-astrometric signals of the Paschen 
$\beta$ lines (tracing the same n=5 level as the Brackett $\alpha$ line used here) from a few T Tauri stars, including DG Tau, which is known to possess a 
strong jet.

\section{Non-Keplerian (radial) flows}
\label{Non_Kepler}
Only some CO spectra of protoplanetary disks are as simple to interpret as the Keplerian
disks discussed above. Many show a structure not consistent with a strictly defined Keplerian velocity field. 
The class of {\it single-peaked} line sources is characterized by a single peaked line spectrum, but with a broad base extending
to $\sim 50\,\rm km\, s^{-1}$ and narrow astrometric spectra that are highly asymmetric about the continuum 
at a characteristic position angle. This spectral class was described by \cite{Najita03}. \cite{Bast10} notes 
that the narrow central peaks in these sources do not show a splitting at least down to the
CRIRES spectral resolution of $\sim 3\,\rm km\,s^{-1}$. 
One obvious explanation for the narrow peak, favored by Occam's razor, involves a Keplerian disk with line emission from large radii with corresponding low Keplerian velocities. 
The spectro-astrometry results rule out this scenario. First, if the lines are formed in a Keplerian disk, the central peak
must be emitted from radii of $R_{\rm peak}\sim GM_*(2\sin{i}/v_{\rm CRIRES}^2)$, corresponding to $>30\,$AU for $i>15\degr$. Such extended emission
should be directly spatially resolved by classical imaging with CRIRES (which has a diffraction-limited 4.7\,$\mu$m spatial resolution of $\sim 20\,$AU at 125\,pc), 
yet no extended line emission is observed. However, classical imaging still allows for the possibility of disks with $i<15\degr$. This is where spectro-astrometry steps in by constraining 
the line emission to much smaller spatial scales ($\lesssim1\,$AU), regardless of inclination, thus ruling out formation in a Keplerian flow at the $\sim$$100\sigma$ level. 

\begin{figure*}
\centering
\includegraphics[width=3.5cm]{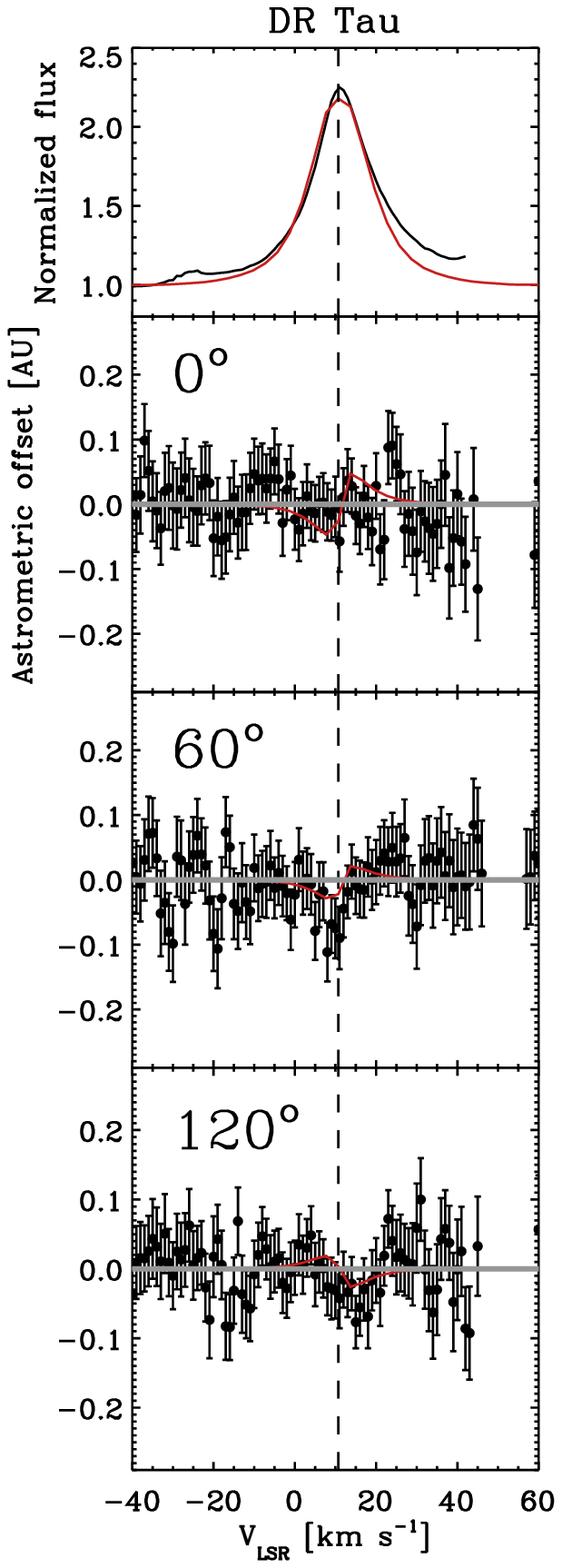}
\includegraphics[width=3.5cm]{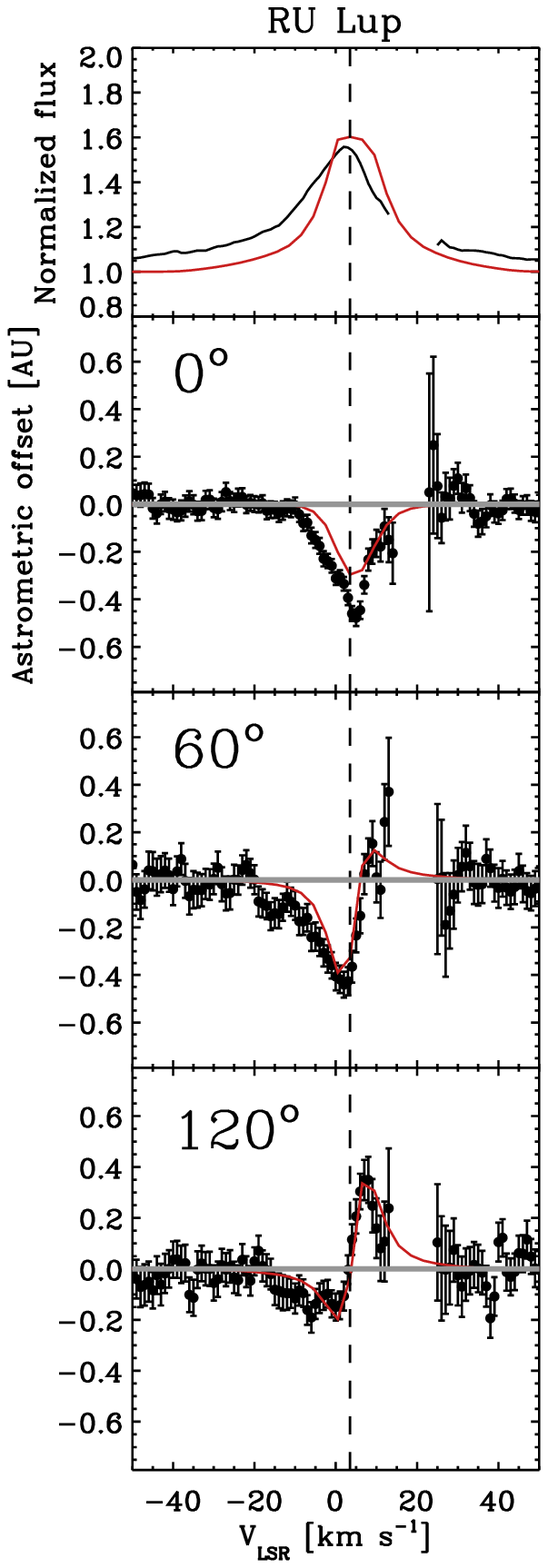}
\includegraphics[width=3.5cm]{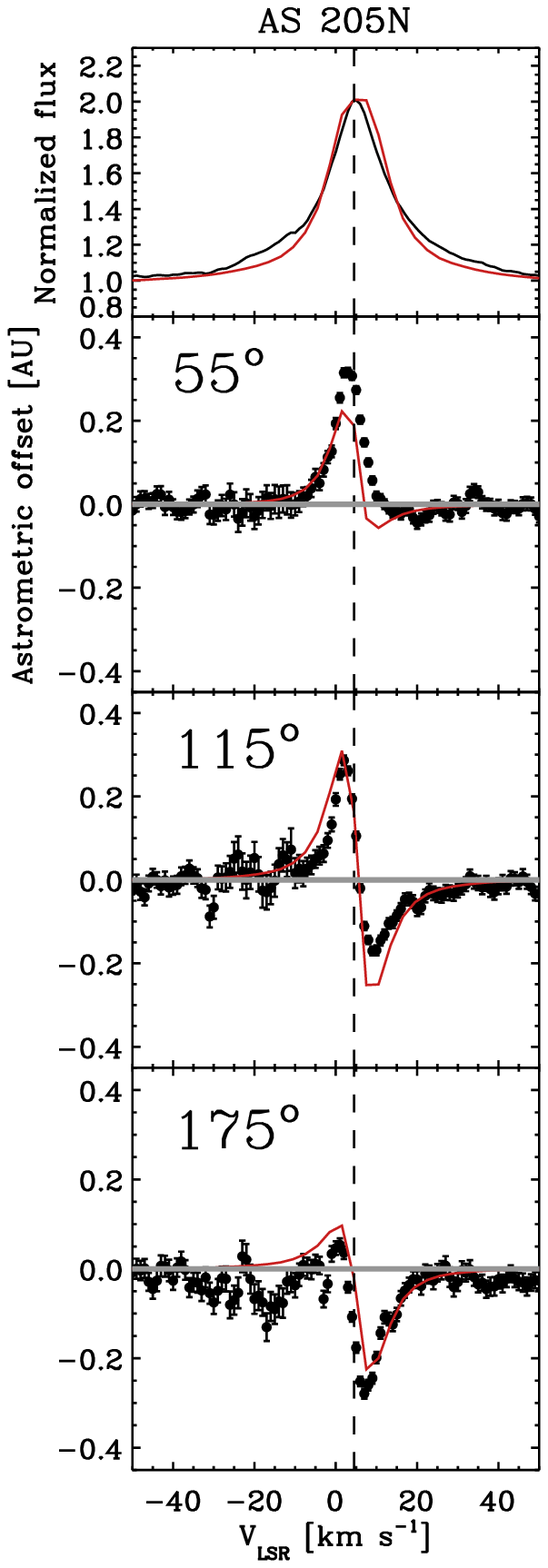}
\includegraphics[width=3.5cm]{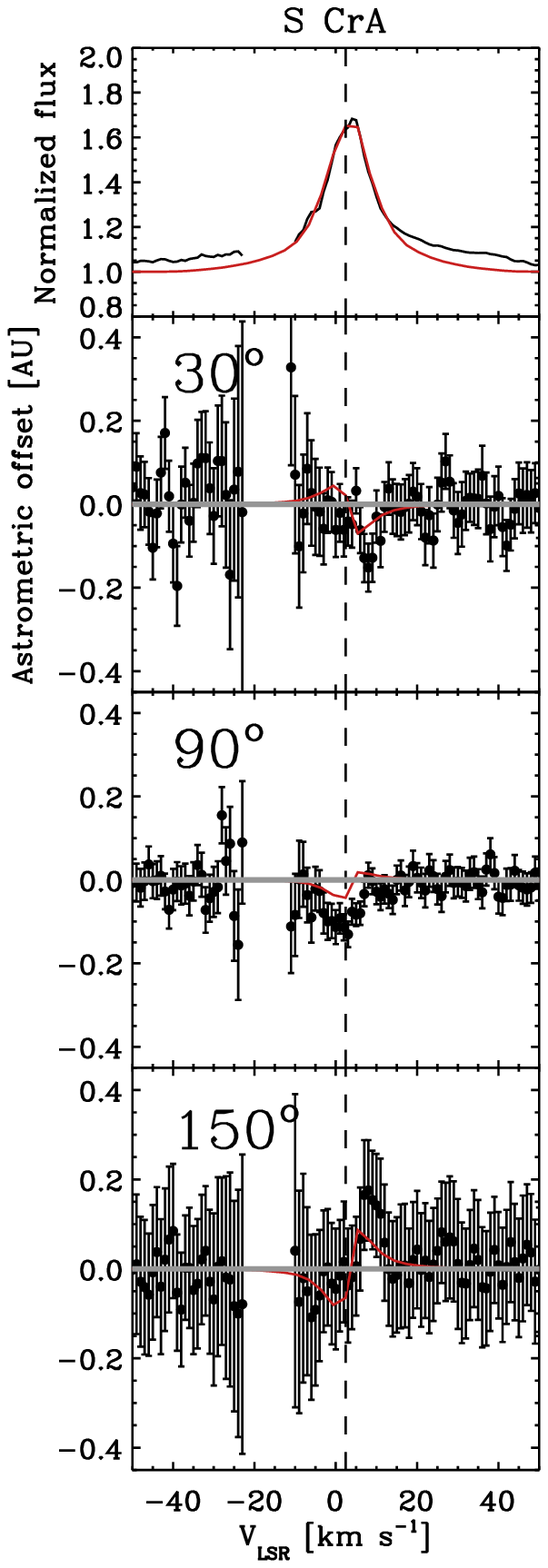}
\caption[]{Flux spectra (top) and spectro-astrometry (lower panels) for the single-peaked line sources. The red curves show possible 
wind models (see Table \ref{wind_table})}
\label{peaky_sources}
\end{figure*}

AS~205N is the prototypical example of a source with single-peaked, broad winged line profile. The 
astrometric spectra of the three clean single-peaked sources are shown in Figure \ref{peaky_sources}. 
For these sources, one out of three position angles, separated by $60\degr$ shows an astrometric line signature 
centered on the line velocity, but entirely offset in one direction. The other two angles show the antisymmetric 
signature that might be expected from a Keplerian velocity field. The amplitude of the astrometric spectra are $\lesssim 1\,$AU, and
the astrometric line is significantly narrower than the flux profile, with the wings missing. 

In essence, the combination of narrow line profiles and small spatial extent indicates sub-Keplerian motions.
We explore a model that explains the qualitative properties of the single-peaked class of CO line spectra in \S\ref{wind_model}.

\subsection{H$_2$O and OH in AS 205N}

A number of other single-peaked line disks are known to show strong emission from
water vapor at 3\,$\mu$m as well as throughout the mid-infrared range. \cite{Salyk08} demonstrated the existence of lines due to gaseous 
H$_2$O and OH in the $3\,\mu$m hot band for AS~205N. The lines
have low contrast relative to the continuum (5-10\%). However, CRIRES is significantly more sensitive at 3\,$\mu$m than at 4.7\,$\mu$m, 
so a detection of an astrometric signal from the water lines may be possible. AS~205N was observed 
in August 2007 with 3 different position angles to match the CO observation. No astrometric signals were detected 
either for H$_2$O or for OH. The upper limits are in both cases $0.2 \times (1+F_c/F_l)\,$mas, corresponding to $\lesssim 0.5$\,AU
at a distance of 125\,pc, when correcting for continuum dilution, or an emitting area of $\lesssim 0.8\,\rm AU^2$. This is consistent with a measured emitting
area of $0.4\,\rm AU^2$ for the 3\,$\mu$m water lines found by \cite{Salyk08}, and is marginally smaller than the radial extent of the CO emission (0.7\,AU). 
Since the excitation energies of the $3\,\mu$m water and OH lines are higher than those of CO, it is likely that the smaller extent can be explained by
an origin in somewhat warmer gas. 

\begin{figure}
\includegraphics[width=8cm]{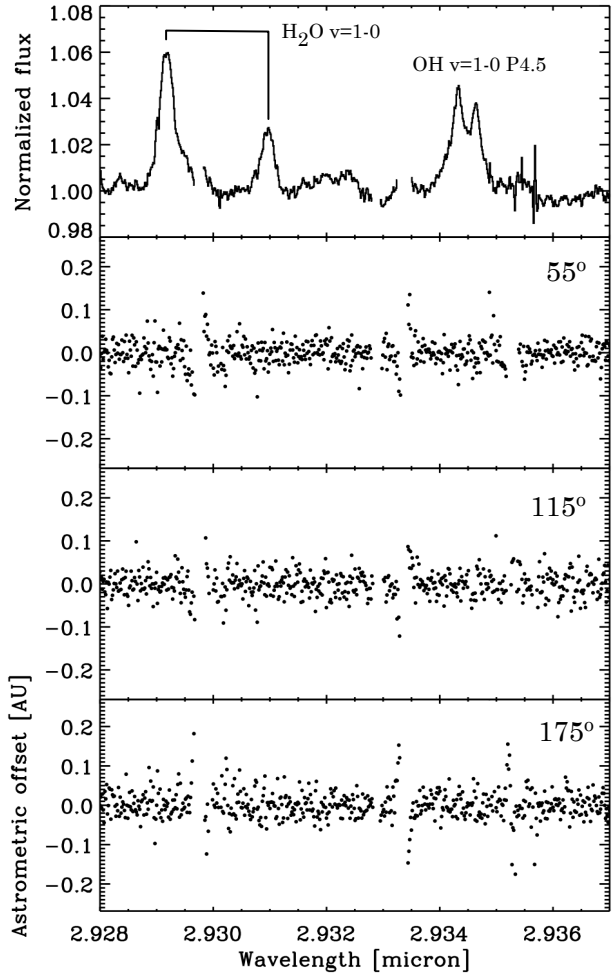}
\vspace{0.4cm} \\
\caption[]{Spectro-astrometry of water and OH lines from the AS 205N disk. }
\label{SA_H2O}
\end{figure}

\subsection{Self-absorbed sources}
Finally, self-absorbed sources have strong absorption components superposed on broad emission lines and are consequently
more complex. While the gas forming the absorption lines in some sources may be completely unrelated to
the disk, it is discussed in section \ref{wind_model} how some self-absorbed sources may be a different representation of
the peaky line sources, namely those viewed at a higher inclination angle.

The self-absorbed sources tend to show the highest amplitude astrometric spectra of the survey. The
physical difference between the single-peaked and self-absorbed sources and the Keplerian disks is
illustrated by their astrometric amplitudes in Figure \ref{size_lum}. These sources do not fall along a neat correlation with the 
stellar luminosity as do the Keplerian disks. T CrA and R CrA in particular show astrometric amplitudes as high as 20-30\,AU, indicating
that the absorbing gas is structured on much larger scales than the emitting gas. These sources are likely to be younger and more embedded than the remaining sample, which
may impact the observables.

\begin{figure*}
\centering
\includegraphics[width=3.5cm]{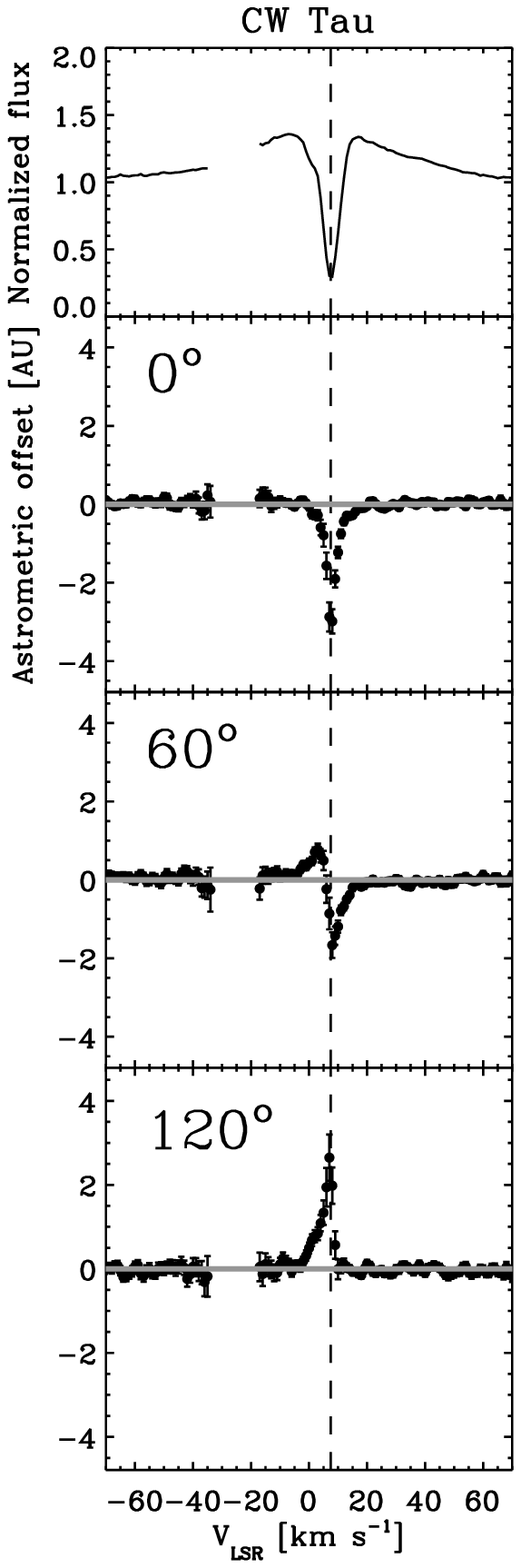}
\includegraphics[width=3.5cm]{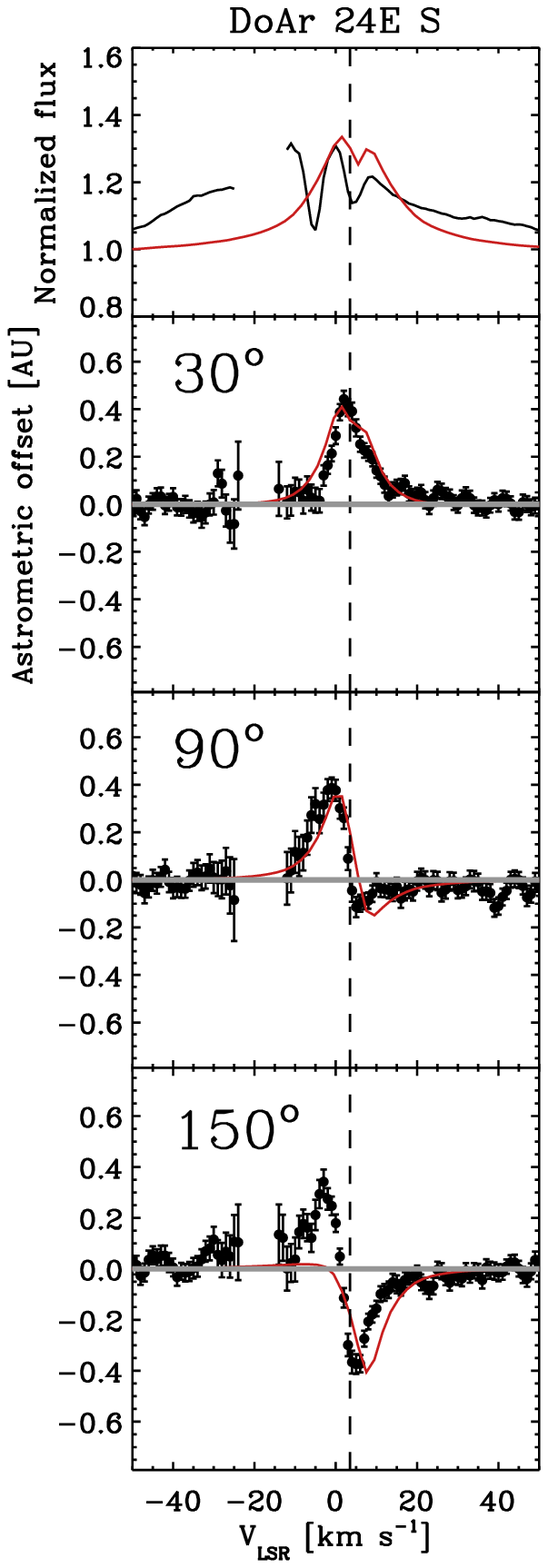}
\includegraphics[width=3.5cm]{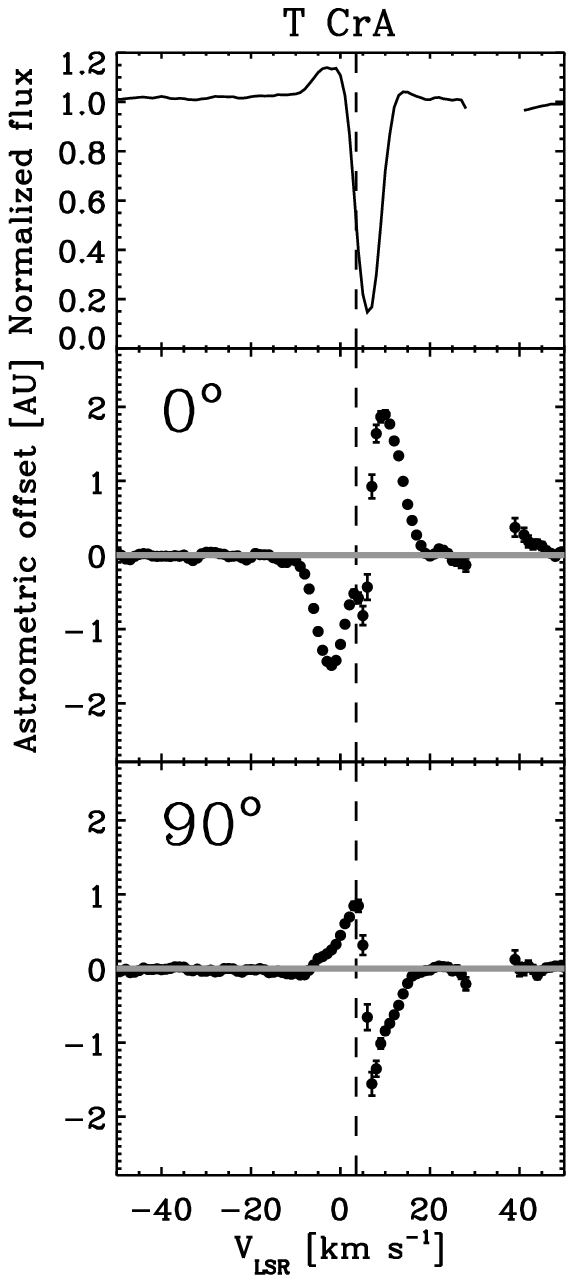}
\includegraphics[width=3.5cm]{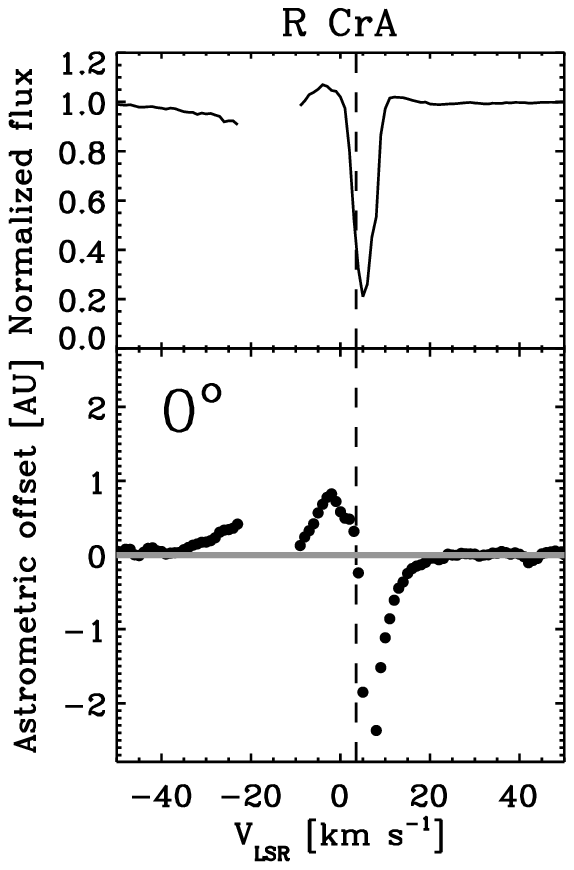}
\caption[]{Spectro-astrometry for the self-absorbed sources. Note that while a model of DoAr24E S exists that fits the spectro-astrometry, the flux line profile is not well matched. 
One explanation for this is that there is an additional compact component to the line. No attempt is made to fit models to T CrA, }
\label{selfabs_sources}
\end{figure*}

\section{A new unified disk+wind model}
\label{wind_model}
\subsection{Parametrization}

\begin{figure*}[ht]
\includegraphics[width=16cm]{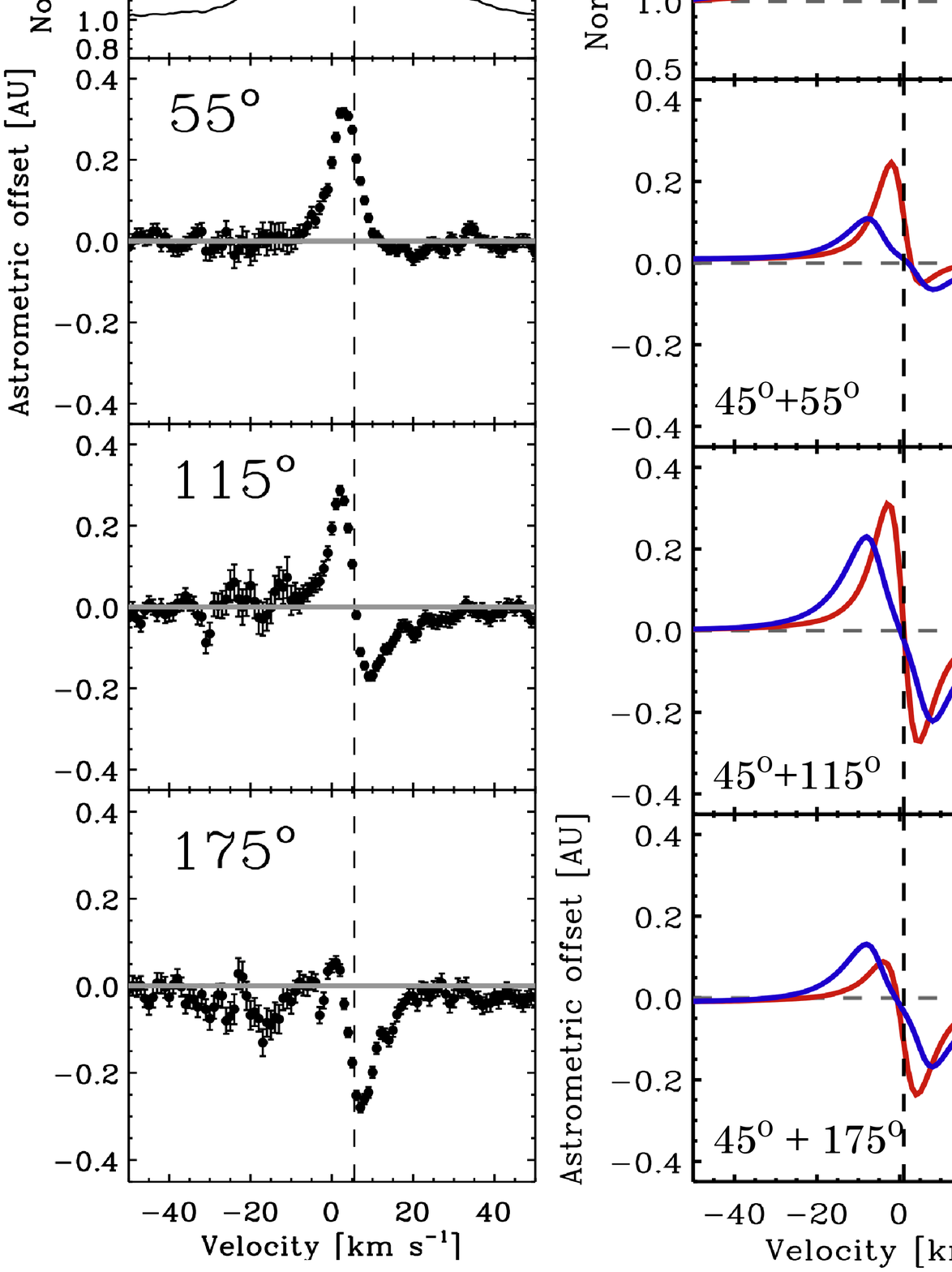}
\caption[]{Sketch of a disk wind model that qualitatively reproduces the broad single-peaked CO line spectra and CRIRES spectro-astrometry from highly accreting T Tauri stars
 -- in this example, AS 205N, a T Tauri star in Ophiuchus. The left panels show the observed CO rovibrational ($v=1-0$ around 4.7\,$\mu$m) line spectrum and
spectro-astrometry at 3 different slit position angles. The middle panels show two 2D radiative transfer models: one for a purely Keplerian disk, and one
for a Keplerian disk in which a wide-angle molecular wind is launched from the disk surface. The right panel contains a sketch of the basic wind geometry.
}
\label{Wind_sketch}
\end{figure*}

\begin{figure}[ht]
\includegraphics[width=8cm]{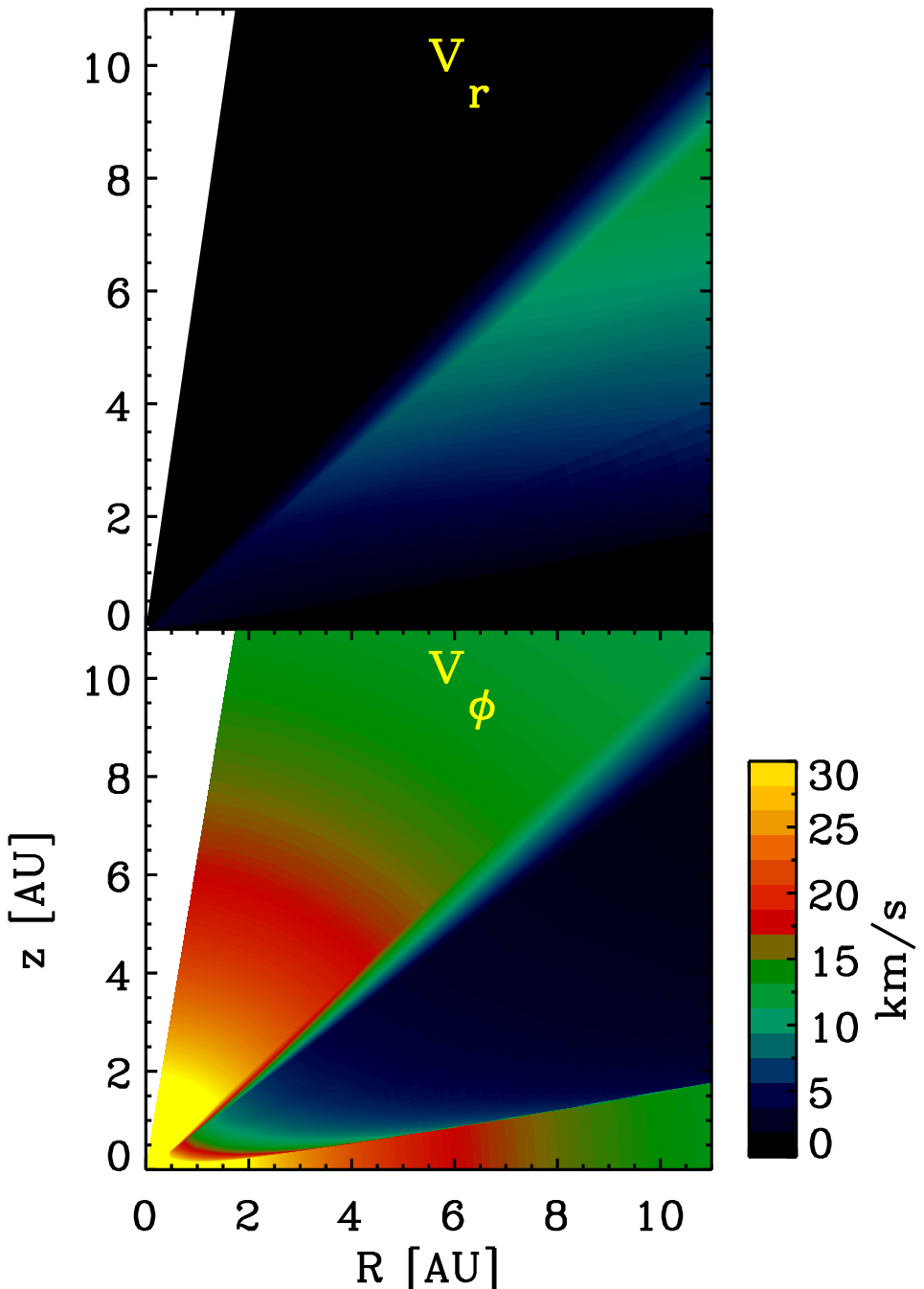}
\caption[]{Radial and azimuthal velocities of the gas in the parametrized disk+wind model. For this set of parameters, which reproduce the observables in, e.g., AS 205N, the wind is seen to
be strongly sub-Keplerian beyond 0.5 AU. Note that the finite radius of the inner rim of the disk and a locus distance $>0$ prevents the wind from filling in all polar angles. In this case, the 
wind only exists at polar angles of $\theta\gtrsim 50\degr$ (as measured from the disk axis). For angles closer to the disk axis, the velocity field is still Keplerian, although the
gas density here is far too low to generate any line emission. }
\label{Wind}
\end{figure}

A parametrized disk wind model is used to test whether an idealized wind structure can produce a plausible match to the phenomenology of the single-peaked line profiles and associated
asymmetric astrometric spectra. It is based on models for UV resonance and hydrogen line emission from accretion disk winds \citep[e.g.,][]{Knigge95, Kurosawa06}, 
and represents a computationally convenient structure inspired by numerical results from magneto-hydrodynamical (MHD) simulations of
magnetized disk winds \citep{Koenigl00}. 

The basic requirement of the observed spectro-astrometry is that the line forming gas must be orbiting the central star with
strongly sub-Keplerian azimuthal velocities in order to produce the single peak without requiring that the emission is extended
at the spatial resolution of CRIRES ($\sim$0\farcs15). A wide angle wind provides a convenient physical 
way of accomplishing this through simple conservation of angular momentum -- as a gas parcel is forced outwards due to the wind pressure, 
the azimuthal velocity decreases linearly with radius, in comparison with the underlying Keplerian disk in which the velocity experiences a shallower decrease as $R^{-1/2}$. 
This generates gas above the disk that is supported by wind pressure and orbits at low azimuthal velocities. 

Following \cite{Kurosawa06}, the wind is constructed as a set of linear streamlines with a locus below the central star at a distance $d$ in units of $R_*$.  This
generates a conical wind with no flow along the disk axis. Briefly, the wind is accelerated along the 
field lines as:

\begin{equation}
v_{\rm stream} = c_s + (f*v_{\rm esc}-c_s) \times (1-A_{\rm scale}/l+A_{\rm scale})^{\beta},
\end{equation}
where $l$ is the coordinate along the stream line, $c_s$ is the sound speed, $v_{\rm esc}$ is the asymptotic velocity at 
the end of the stream line and $A_{\rm scale}$ is the scale of the acceleration region of the wind. $\beta$ is the
wind acceleration parameter. Requiring angular momentum conservation, the azimuthal velocity component is:

\begin{equation}
v_{\phi} = v_{\rm Kepler}(l_0) \times \frac{R(l_0)}{R},
\end{equation}
where $R$ is the radial disk coordinate. 

The density of the wind is calculated assuming mass conservation:
\begin{equation}
\rho_{\rm wind} = \frac{\dot{\Sigma}(R)}{v_{\rm stream}|\cos\delta|}\times\left[\frac{d}{S\cos\delta}\right]^2
\end{equation}
Here, $\dot{\Sigma}(w) \propto R^{-p}$ is the local mass-loss rate, $\delta$ is the angle between the stream line and the disk normal and S is the distance to the wind locus. 
The exponent of the local mass loss rate is taken to be $p=7/2$ \citep{Krasnopolsky03}. The
total wind mass loss rate can be calculated by integrating over the disk and multiplying by two to include the opposite surface:

\begin{equation}
\dot{M} = 2\int_{R_{\rm inner}}^{R_{\rm outer}} 2\pi R \dot{\Sigma}(R) dR
\end{equation}

\begin{figure}
\centering
\includegraphics[width=8cm]{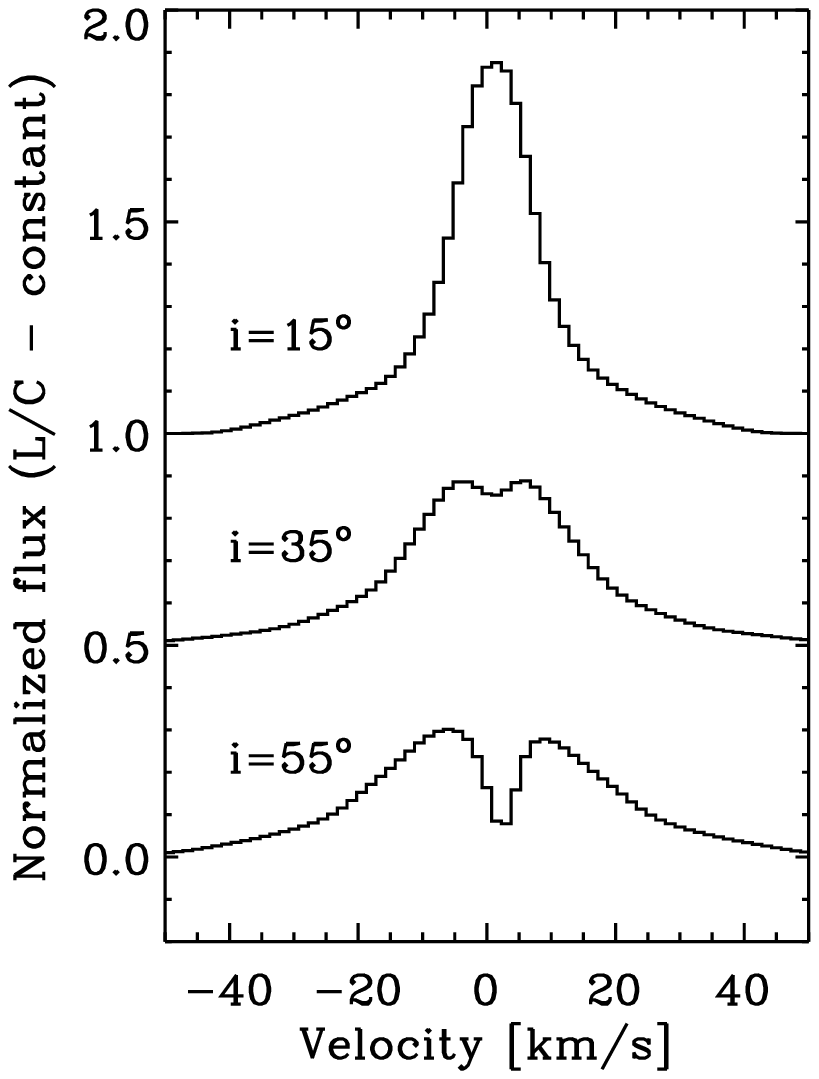}
\caption[]{Effect of inclination on the line profile in the disk wind model. }
\label{peaky_incl}
\end{figure}

The raytracer RADLite \citep{Pontoppidan09} is used to render model lines and spectro-astrometry for the wind models, based on a generic
model of a flared protoplanetary disk, and assuming level populations in LTE. Specifically, the temperature structure is assumed to be in
equilibrium with the stellar radiation field and dominated by dust heating/cooling. In reality, the heating of the wind is likely to be dominated by 
photo-electric heating similar to the heating of the disk atmosphere \citep{Kamp04, Jonkheid04, Gorti04, Dullemond07} or, perhaps, ambipolar diffusion \citep{Safier93}. The cooling
may be dominated by adiabatic expansion and molecular cooling (e.g., partly via the observed CO and H$_2$O lines). However, we restrict ourselves
to qualitative models in this paper (see also \S\ref{Caveats}), since a detailed and appropriate physical treatment of the thermal wind structure required to
match the observations will be likely be a significant study in its own right. 

\subsection{Properties of the wind model observables}

Figure \ref{Wind_sketch} illustrates the wind geometry and compares the observables generated using the wind model for
the spectro-astrometry of AS~205N. The total mass-loss rate is $9\times 10^{-9}\,\rm M_{\odot}\,yr^{-1}$, assuming a CO abundance
of $5\times 10^{-5}$ relative to $\rm H$. This mass-loss rate is consistent with a 
highly accreting CTTS. Indeed, AS~205N is accreting at a rate of $7\times 10^{-7}\,\rm M_{\odot}\,yr^{-1}$ \citep{Eisner05}. There is 
no treatment of the chemistry of the wind in the present model, so if the abundance of CO is lower than that assumed, the mass-loss rate will be correspondingly higher. 
However, the numbers appear to be physically reasonable; the ratio of mass-loss to mass-accretion rates is $\dot{M}_{\rm wind}/\dot{M}_{\rm acc}\gtrsim 1.5\%$, 
and as this ratio is expected to be as high as 10\% \citep{Koenigl00}, the CO abundance in the wind could in fact be lower than assumed. 
There are many free parameters in the wind model, so that shown represents one possibility
that reproduces the main characteristics of the observed astrometric spectra. A full parameter study, beyond the scope of 
this paper, may reveal significant degeneracies. That said, some parameters are well constrained, including the position angle of the system. 
General properties are also locus distances $d\lesssim 10\,R_*$ and inclinations of $i\lesssim 30\degr$. The velocity field
used for AS~205N is shown in Figure \ref{Wind}. 

It is seen that the wind model can explain the key properties of the combined line profile and astrometric spectra:
namely it readily produces single-peaked lines with broad wings. The wings are dominated by the innermost, Keplerian disk, 
while the single peak is generated by the sub-Keplerian gas at a few AU. The presence of a strongly
sub-Keplerian component requires a wide-angle, non-collimated wind, in the model parametrized as a small
star-to-locus distance, $d$.  Increasing $d$ results in a double-peaked line profile that would
be resolvable with CRIRES. The astrometric spectra become highly asymmetric
at P.A.'s close to the disk minor axis. It is important to note that this wind geometry does not necessarily produce strongly blue-shifted lines -- 
which might have been expected. The uncollimated winds sees much of the line emission coming from gas moving on trajectories nearly parallel to the disk surface. 
For sources viewed at low inclination angles, this generates low radial velocities and prevents large blue-shifts, as
required by the data. The addition of a significant amount of gas at high altitudes where the optical depths toward the central star are low, results in
higher column densities of warm CO, leading to brighter lines with higher line-to-continuum ratios, a property of this class of CO spectra
that has been observed \citep{Bast10}. 

Table \ref{wind_table} summarizes the RADLite wind model parameters for the wind-dominated disks. Since we do not 
fully explore the parameter space, the entries in the table represent a possible model, but with a caution that significant
degeneracies may exist. Further, the use of LTE level populations and the assumption of coupled dust and gas temperatures 
almost certainly introduce inaccuracies in the implied wind structures. Indeed, it was typically necessary to increase the luminosity of the central
source to values well above the known stellar properties to match the amplitude
of the astrometric signal, especially in the case of RU Lup, the reason being that the gas temperatures are significantly higher
than what can be explained by pure coupling to dust. 

\subsection{The wind model at higher inclinations as a model for the self-absorbed disks}
As can be seen in Figure \ref{Wind}, there is no wind component along the disk axis due to the finite size of the inner disk rim. Because of this, 
a disk wind viewed close to face-on, as is likely the case for AS 205N, allows a free view of the continuum emission from the innermost disk, consistent
with the low extinction toward the central star in many of our targets. 
However, if the inclination were higher, say $\sim 45\degr$, the entire wind column will be located between the disk continuum emission and the
observer. This leads to the formation of a strong line absorption component, as illustrated in Figure \ref{peaky_incl}. The Figure shows the line profiles of the prototypical
AS 205N model, but viewed at higher inclination angles. A deep absorption line forms at inclinations higher than $\sim 40\degr$ for this particular model, a result that can
be compared directly to sources exhibiting a self-absorbed line, such as CW Tau. The line profile of this source is qualitatively reproduced by
the AS 205N model viewed at an inclination angle of $55\degr$. The depth of the absorption line is smaller in the model, but can easily be deepened by
increasing the mass-loss rate. 
We note that the small spectro-astrometric sample of self-absorbed source includes 
sources such as R CrA and T CrA are clearly younger and more embedded than the remaining disk sample, and their
deeper absorption components are consistent with mass-loss rates 2-3 orders of magnitudes higher than that of the AS 205N model. 
However, there are more bona-fide isolated disks in our CRIRES survey showing strong, warm CO absorption that were not included as spectro-astrometric targets, 
but with properties otherwise resembling CW Tau. 

The four self-absorbed disks are not modeled in detail for several reasons. We found it difficult to reproduce the very high amplitudes of
the astrometric offsets for the absorption component for CW Tau and T CrA, likely indicating the presence of extended continuum emission structure. 
DoAr 24E S has several absorption components that also cannot be modeled exactly with the wind model without the additional parameters. 
However, we do still present a best effort model of DoAr 24E S in Figure \ref{selfabs_sources} to illustrate the required complexity of this star.

\begin{table}
\centering
\caption{Model parameters for wind-dominated sources}
\begin{tabular}{llllll}
\hline
\hline
Star  & P.A.\tablenotemark{a} & i\tablenotemark{b} & $L_{\rm model}/L_*$\tablenotemark{c} & d\tablenotemark{d} &$\dot{M}_{\rm wind}$\tablenotemark{e} \\
        &        &   &                              & $R_*$ & $M_{\odot}\,\rm yr^{-1}$ \\
\hline
DR Tau         & 0\degr:    & -9\degr & 1.0 & 1 & $4\,10^{-8}$\\
RU Lup         & 80\degr   & 35\degr & 90 & 8  & $4\,10^{-9}$\\
AS 205N       & 235\degr & 20\degr & 1.8 & 5  & $9\,10^{-9}$ \\
DoAr24E S    & 235\degr & 20\degr & 1.8 & 5  & $9\,10^{-9}$ \\
S CrA            & 15\degr  &  10\degr & 1.3& 4  & $1\,10^{-8}$ \\
\hline
\end{tabular}
\tablenotetext{a}{Position angle of the disk major axis.}
\tablenotetext{b}{Inclination of the disk rotation axis. A negative value of the inclination corresponds to the north pole of the disk facing the observer. }
\tablenotetext{c}{Ratio between the luminosity of the central source used in the model and the actual stellar luminosity. A high ratio
indicates that the gas is heated to temperatures significantly higher than those of the dust.}
\tablenotetext{d}{Distance between the locus of the wind and the central star (in units of stellar radius). d=0 corresponds to a spherical wind.}
\tablenotetext{e}{Mass loss rate of the wind assuming $x({\rm CO})=5\,10^{-5}$.}
\label{wind_table}
\end{table}

\section{Discussion}
\label{discussion}
The origin of CO rovibrational emission lines from protoplanetary disks has long been a puzzle. While it has been clear that
the lines are formed close to the star, given the high temperatures and densities required to excite the lines, the great variation 
of the line shapes and excitation temperatures -- rotational and vibrational -- have evaded a unified explanation. Specifically, the
lines cannot be explained solely by thermal emission from simple Keplerian disks, except in a few cases. This was already
noted by \cite{Najita03}, but an unambiguous identification of the additional (to a Keplerian disk) radial flow pattern could not be determined. Now, with the addition of
spatial information, as offered by high resolution spectro-astrometry, we can present a more comprehensive picture of the dynamics of molecular gas 
on AU-scales in protoplanetary disks. 

\subsection{Size of the line emitting regions}
In \S\ref{Kepler_sources} we show that sources with double-peaked (Keplerian) profiles obey a size-luminosity relation with a power-law index of $\alpha=0.5$, as expected for emission
truncated at a radius of constant temperature. Critical ingredients in identifying this relation are the ability to distinguish Keplerian disk-dominated lines from the 
wind-dominated and self-absorbed lines, as well as the use of the direct size measure offered by spectro-astrometry. Specifically, the astrometric amplitude is not
dependent on disk inclination, which otherwise makes it difficult to use the line width as an accurate proxy for size. 

Another conclusion that can be drawn from the measured sizes of the emitting region from Keplerian sources is that a line origin in the accretion flow can be ruled out. Accretion flows are expected to generated double-peaked CO line profiles \citep{Najita03}. Funnel (along
magnetic field lines) accretion flows are expected to be launched from radii near, or within, the corotation radius \citep{Shu94}. This is, for T Tauri stars, located inside the dust sublimation radius at 2-10 stellar radii, and thus well within the observed CO radii of $\sim$100\,$R_*$. The Br$\alpha$ lines as observed in RNO 90 and HD 142527
have astrometric displacements much smaller than those of CO and evidently traces a different gas component, which may still be an accretion flow. Note that because the H~I lines trace a much wider range of infall velocities than the CO -- H~I line widths are 100-200\,$\rm km\,s^{-1}$ -- they are less affected by Keplerian rotation at the bottom of the flows than
the narrower CO lines \citep{Muzerolle98}. Consequently, they may not display a double-peaked profile even if formed as part of an accretion flow.

Is it expected that CO follows a relation that was developed for dust sublimation? One possibility is that dust shielding plays a significant role 
in the survival of CO in the inner regions of disks. In this case, CO would only be able to survive at radii at which the radial column of dust is sufficiently high. This
is consistent with CO existing at radii larger than those of the inner dust rim. However, CO is also know to efficiently self-shield, even at low column densities of 
$N(H_2)\sim 10^{18}\,\rm cm^{-2}$ \citep{Visser09}, although uncertainties are large at temperatures higher than a few 100\,K.

\subsection{Non-Keplerian motions and a unifying model for CO rovibrational lines}
We propose that most, if not all, disks have a molecular wind velocity component in addition to pure Keplerian rotation -- one that is effectively traced by
ro-vibrational CO lines, presumably in addition to lines from a wide range of other molecules with strong infrared emission bands. \cite{Najita03} argues against a wind based on
the fact that the lines are not blue-shifted and that very young (stage I) high accretion rate sources sometimes do not show rovibrational CO emission lines. 
However, our wind model shows that a strong blueshift of the line emission is not necessary for the uncollimated slow disk winds constrained by angular momentum conservation. 
While we do not specifically address the lack of strong CO lines toward some embedded young stellar objects, strong accretion may heat the disk mid-planes
to a point where the temperature inversion in the disk surface required for line emission is no longer possible. Furthermore, 
we do find that the contribution of outflowing gas to the line emission is highly variable, possibly in relation to the level of accretion activity onto the central star. 
While we have not sought to determine the wind launching mechanism, the data do suggest that the wind is uncollimated and slow ($|v_{\rm wind}|<v_{\rm Kepler}$). It is important to stress
that the wind should not be seen as entirely separate from the Keplerian disk, but rather as a modification.  For instance, the 
gas does not suddenly stop its Keplerian rotation even as it attains a radial velocity component. For low wind accelerations, it
may be difficult to determine that there is a difference at all from a pure Keplerian velocity field. As the wind is launched, the gas maintains its angular momentum 
and slows down so much as it is pushed outwards that it may never escape the system, but re-accrete onto the disk at larger radii. 
The mass-loss rates implied by our wind models are large enough that the inner disk gas can be completely redistributed during the lifetime of the disk. This may
limit the time available for planet formation, in line with ideas generally associated with photo-evaporative disk winds \citep{Alexander06}.  

\subsection{Wind launching mechanism}
We will not discuss in detail how the molecular wind is launched. However, the requirement that it is uncollimated and slow likely
places significant physical constraints on the launching mechanism. For instance, a magnetic centrifugal wind might be expected to generate winds
that are much too fast as the gas in this case will be locked to the stellar rotation, known to be $v_{*}>v_{\rm Kepler}$ beyond the inner rim of the disk. 
A thermal wind is much slower, but may require high ionization fractions and gas temperatures ($T_{\rm gas}\sim 10,000\,$K), which 
could be inconsistent with the presence of abundant molecules with rotational temperatures of $\sim 1,000\,$K. The photo-evaporative wind models by 
\cite{Alexander06} predict wind launch speeds that may even be slightly subsonic (5-10\,$\rm km\,s^{-1}$), consistent with 
our observational finding. It can be noted that the evaporative winds predict a somewhat shallower exponent $p$ for the drop-off of the mass-loss rate $\dot{\Sigma}(w)$ with radius than the $p=7/2$
used here. Experimenting with our model indicates that good fits may also be found with $p=5/2$, although not as readily. Also noteworthy is that the models by \cite{Alexander06} predict significant blueshifts of wind-dominated [Ne~II] line peaks \citep[5-10\,$\rm km\,s^{-1}$][]{Alexander08}, which does not
appear in the CO data, at least compared to the ambient cloud velocity traced by cold absorption components.

\subsection{Implications for radial mixing} 
The observational implication of slow gas flows above the canonical ``warm molecular layer'' of the disk may have important implications for 
the transport of material in the disk. Specifically, the gas may never be accelerated to velocities allowing it to escape the disk, allowing it to fall back onto the disk at larger radii. 
For flow rates of $10^{-9}\,M_{\odot}\rm \,yr^{-1}$, a wind lasting a few Myr will clearly be able to re-distribute a significant fraction, if not all, of the inner disk gas not accreted onto the central star.  Similar ideas
where driving the development of early disk wind models, such as the X-wind \citep{Shu94} to explain the redistribution
of solids required to produce the crystalline dust grains in cometary material. It may be important to note that if the outflowing gas falls back onto the disk, there will be
a significant azimuthal velocity differential as sub-Keplerian material impacts the Keplerian disk below. Judging from Figure \ref{Wind}, the differential may be as high as 10-15\,$\rm km\,s^{-1}$
if the fall happens at 10\,AU, but less if the fall happens farther out in the disk. This is sufficient to generate shocks that may be 
observable. 

\subsection{Caveats, unknowns and new questions}
\label{Caveats}
Clearly, the parametrized wind model is not based on a detailed treatment of the underlying hydrodynamics and radiative transfer. However, it
does represent a phenomenological structure required to match clearly defined observables. It is therefore important to consider 
the circumstances under which the physical and chemical structure of the wind may differ from the simple model. 
 
One property of the wind model that does not match the observations is the rotational temperatures of the lines. The model predicts lower
temperatures than observed. However, this can likely be explained by the use of LTE level populations in thermal balance with the dust. The gas that forms the wind will be exposed to a strong IR and UV radiation field from the central star and innermost disk, and is therefore subject to fluorescence pumping as well as
collisional excitation by a gas that is heated and likely partly ionized by photo-electrons. Balancing this are the strong cooling terms offered by
the rovibrational molecular emission, in particular that of CO and water. While previous wind models \citep[e.g.,][]{Safier93} generally
predict very high wind temperatures (10,000 K) and fast winds ($\rm 100\,km\,s^{-1}$), the existence of significant molecular coolants may, in part of the wind flow, maintain
the required low temperatures and velocities. In particular the temperatures may be closer to those given by the 
assumption of equilibrium with the stellar radiation field, as assumed in this paper, than the high temperatures implied by a purely
atomic gas. A future detailed heating-cooling balance calculation for the wind model is clearly
an important next step.

More detailed hydro-dynamic modeling is essential to fully assess the implications of slow disk winds. What is the wind launching mechanism? 
Does the wind eventually fall back onto the disk and at which radii? How can the wind remain, at least in part, in a molecular form as the material is lofted to altitudes where
dust shielding low and the gas is exposed to a harsh UV radiation field? While a model of the chemistry of the molecular wind is a study in its own right, it can be noted that the
CO column densities required to generate the deep absorption lines seen in the self-absorbed (more inclined) sources are likely high enough to self-shield. Likewise, the
presence of water in the flow, at least in AS 205N -- see Figure \ref{SA_H2O} -- will provide additional shielding against UV photons for a range of molecular species \citep{Bethell09}. 

\section{Conclusion}
\label{conclusion}
Using spectro-astrometry to image molecular gas at 0.1-10\,AU in a sample of protoplanetary disks has revealed an intriguing range of structure. 
Some disks appear to be dominated by gas in Keplerian rotation about the central star, as expected, while others have a significant {\it slow} radial velocity component consistent with
a wide-angle disk wind. The basic observational evidence for a slow wind is the clear presence of {\it low velocity gas ($\sim 5\rm\,km\,s^{-1}$) within a few AU from the central star.}
It is difficult to generate low velocity gas deep in a potential well, but one simple way of doing so is via angular momentum conservation of an expanding, but initially Keplerian, flow. 
While it has long been known that fast atomic winds were common in T Tauri stars, we now find that there is a significant molecular component as well. Further, 
it appears that the observed molecular wind must be launched from the disk surface. It is not clear whether the gas in the wind reaches escape velocities, and
may therefore be re-accreted onto the disk at larger radii. The wind reproducing the CO line spectro-astrometry is much slower than that 
predicted by X-wind theory, which reaches terminal velocities of several $100\,\rm km\,s^{-1}$ \citep{Shu94}, but is a much better match to photo-evaporative
flows that have poloidal velocities similar to the sound speed \citep[e.g.,][]{Alexander06}. The existence of disk winds with high mass-loss rates have significant implications for the
availability of material for planet formation in the PFZs of protoplanetary disks, and may limit the time available for planet formation. The potential large scale cycling of inner disk material as suggested by the low velocity of the winds will also influence the chemistry of planet-forming material, for instance by exposing a large fraction of the disk mass to the strong UV fields at high disk elevations. Future work will include
the development of a model for how slow {\it molecular} winds are launched, as well as a chemical model of the wind that can explain the survivability of, 
at least, CO, H$_2$O and OH at low optical depths and at high elevations above the disk.

\acknowledgments{The authors are grateful to Colette Salyk and Ewine van Dishoeck for comments that improved the manuscript. Joanna Brown and Bill Dent are thanked for
obtaining some of our spectro-astrometric data. A very special thanks goes out to all the Paranal personnel that assisted with
our visitor observations, without whom this study would not have been possible. Support for KMP was provided by NASA through Hubble Fellowship grant \#01201.01 
awarded by the Space Telescope Science Institute, which is operated by the Association of 
Universities for Research in Astronomy, Inc., for NASA, under contract NAS 5-26555. This paper is based on observations made with ESO Telescopes at the Paranal Observatory 
under program ID 179.C-0151.
}

\bibliographystyle{apj}
\bibliography{ms}

\appendix
\section{Notes on individual sources}
\subsection{Keplerian disks}

\paragraph{LkHa~330}
This transitional disks is one of the ``cold disks'' imaged at submillimeter wavelengths by \cite{Brown09}. The spectro-astrometry reveals a significantly smaller inclination angle (12$\degr$) than that suggested by the ellipticity of the 850\,$\mu$m continuum image (42$\degr$). The submillimeter image also shows that dust has been depleted within $\sim$50\,AU, while we find that molecular gas persists at 4\,AU, but not closer to the star. In many ways, this disk appears to be similar to that of another transition disk, SR 21, 
described in \cite{Pontoppidan08}. 

\paragraph{GQ~Lup}
This star is probably best known for having a substellar companion, GQ~Lup b \citep{Neuhauser05}, currently located due west of GQ~Lup. 
The spectro-astrometry of the GQ~Lup disk shows that it is oriented along the N-S axis and has a relatively high inclination of $65\pm 10\degr$. 
It is not known whether GQ~Lup b orbits in the disk plane, but if it does, the physical separation will be higher than the projected separation of 0\farcs7 
by a factor 2.4, bringing the physical separation to 240 AU. Further, the prediction is that any orbital motion of the companion will be along the N-S axis.
It can be noted that the stellar inclination of GQ~Lup has been found to be $27\pm 5 \degr$ \citep{Broeg07} -- very different than that of the disk, raising the question of
whether this difference is due to interactions between the companion and the disk. 

\paragraph{HD~142527}
The disk around this borderline Herbig Ae/Be -- T Tauri star (Spectral Type = F6) has been imaged in the continuum at near- to mid-infrared wavelengths. At mid-infrared
wavelengths (18-25\,$\mu$m), the outer disk exhibits a strongly asymmetric structure, with the eastern side being much brighter than the western \citep{Fujiwara06}. 
These authors suggested that the bright eastern arc being an illuminated rim of a disk inclined along a north-south axis. The spectro-astrometry, however, demonstrate
that the actual axis of the (inner) disk is at a P.A.=293\degr. Near-infrared, scattered-light imaging by \cite{Fukagawa06} find that the disk is elliptical along a major axis
in the NW-SE direction, more consistent with the spectro-astrometric P.A. Further, at NIR wavelengths, the south-western part of the disk is brighter, indicating that
this is the closer part of the disk, exhibiting strong forward scattering. The inclination angle of $25\degr$ matches both the mid-infrared and near-infrared imaging. 

\paragraph{HD~144432}
The polarization axis of the disk around HD 144432 was found to be
114\degr \citep{Rodrigues09}. \cite{Monnier06} determined the size of
the H-band source (tracing an inner rim of the disk) to 0.5\,AU using
closure-phase interferometry, or slightly smaller than the CO emitting region. 
 
\paragraph{RNO~90}
While this G5 T Tauri star has featured prominently in studies of mid-infrared spectroscopic disk properties \citep{Kessler06, Pontoppidan10}, little is known
about the geometry of the disk. Using CO spectro-astrometry, it can be determined that the inclination is $37\pm 5\degr$ and that the major axis of the 
disk is oriented along the N-S axis with a P.A. = $177\pm 3\degr$. There is no apparent inner gap in the CO emission, which can be traced to within 0.1\,AU. 

\paragraph{VV~Ser}  
This Herbig Ae/Be star is known to have a prominent disk shadow \citep{Hodapp04, Pontoppidan05}. The disk orientation, P.A. and inclination was determined by
\cite{Pontoppidan07a, Pontoppidan07b} using the disk shadow in combination with near-infrared interferometric visibilities. The P.A. 
of \cite{Pontoppidan07b} of $15\pm 5$ can be compared to the spectro-astrometric P.A. of $17\pm 5\degr$, while the disk shadow inclination of $70\pm 5\degr$
can be compared to the astrometric inclination of $65\pm 5\degr$. It is important to note that the likely distance to the Serpens
star forming cloud, of which VV Ser is a member, has been significantly revised from 260\,pc \citep{Straizys96} to 410\,pc \citep{Dzib10}. The radius of the 
K-band continuum emission determined by \cite{Pontoppidan07a} is then revised to 1.1\,AU, in comparison to a CO emitting radius of $3.4\pm 0.4\,$AU. 

\subsubsection{Single-peaked line sources}

\paragraph{DR~Tau} 
DR~Tau is a well-known high activity T Tauri star with a high degree of optical veiling \citep{Hessman97} and spectral variability \citep{Smith97, Alencar01}.
Near-infrared interferometry has provided an uncertain measure of the disk P.A.=$160\pm 55\degr$ \citep{Akeson05}, consistent with our spectro-astrometry.
\cite{Andrews07} finds a disk P.A.$=170\pm 8\degr$, while \cite{Isella09} finds a P.A.=$98$ and an inclination of 37\degr, using high resolution (sub)millimeter interferometry. 
Based on stellar $vsini$ measurements coupled with periodic variability, \cite{Muzerolle03} finds a high stellar inclination of 69\degr. In summary, the
basic geometry of the DR Tau disk is still uncertain, but the spectro-astrometry indicates an uncertain P.A. close to 0\degr and a low inclination. 

\paragraph{RU~Lup}
RU Lup is a well-known classical T Tauri star with strong accretion signatures. \cite{Gahm05} suggested that
the star may have a close stellar companion due to the presence of periodical radial velocity changes, although the presence of
starspots leading to the observed radial velocity signal is not ruled out (see discussion in \cite{Herczeg05}). If the
variability is due to starspots, a measurement of $v\sin{i} = 9\pm 1\,\rm km\,s^{-1}$ \citep{Stempels02} implies a nearly face-on
orientation of the star, and presumably the disk as well, of $i=24\degr$ \citep{Herczeg05}. \cite{Takami01} presented spectro-astrometry
of the $H\alpha$ line and found evidence for a jet with a P.A.$\sim 45\degr$. The CO P.A. of $80\degr$ is significantly different. 

\paragraph{AS~205N} 
This star, also known as AS~205A, the primary of the close binary (separation 1\farcs3) system known for exhibiting bright molecular emission from water and organics throughout the mid-infrared \citep{Salyk08}. 
Since the CRIRES PSF has a width of $\sim$0\farcs12, there is no contamination from the secondary. The central star has a late spectral type (K5) relative to its luminosity (4\,$L_{\odot}$), 
resulting in a very young apparent age of $5~10^{5}\,\rm yr$ \citep{Andrews09}. A comparison to the disk geometry of \cite{Andrews09}, derived with high spatial resolution
submillimeter interferometric continuum and line imaging with the SubMillimeter Array (SMA), is instructive. They find, using the dust continuum contours, an inclination of $\sim 25\degr$, which is
consistent with the wind model requiring the disk to be nearly face-on. Their P.A. is different from that derived from the spectro-astrometry. While continuum isocontour fits
can be uncertain for nearly face-on disks, spatially resolved SMA CO (3-2) images clearly show rotation consistent with a disk at P.A.$\sim$165\degr. In contrast, 
the spectro-astrometry requires a P.A.$\sim 235\degr$ of the AS~205N disk major axis. We suggest that the discrepancy could be explained if the rotation pattern seen in the rotational CO line is due
to contamination from AS~205S; a scenario which is explicitly not ruled out by \cite{Andrews09}. If so, this implies an interesting configuration in which the orbit of the AS~205 binary {\it is not
coplanar} with the AS~205N disk. Non-coplanarity has been observed as a general property of binaries with disks, although there is probably some alignment for disks around stars with
separations between 200-1000\,AU \citep{Jensen04}. 

\paragraph{S~CrA N} S~CrA is a $1.3\arcsec$ binary. Our spectro-astrometry is of the northern component, the primary at 
near-infrared wavelengths, but the secondary in the visible. There is, to our knowledge, no other measurements of the
disk geometry of this source in the literature. 

\subsubsection{Self-absorbed sources}
\paragraph{CW~Tau}
CW Tau has an optical jet with a P.A.$=155\degr$ and inclination of $41\degr$ \citep{Gomez93, Hartigan04, Coffey08}, qualitatively 
consistent with the CO spectro-astrometry.
\paragraph{DoAr~24E S}
This is a $2\arcsec$ binary, with the south-eastern component being an InfraRed Companion (IRC) \citep{Chelli88}. Our spectro-astrometry is
of the secondary (at K-band) southern component - this is the brighter, by about 0.7 magnitudes, component in the M-band.
The secondary disk has P.A. of 11\degr as determined using K-band polarization \citep{Jensen04}. The secondary is itself a possible 
close, equal brightness, binary with separation 5.6\,mas and P.A.=44\degr \citep{Koresko02}, based on speckle imaging. The presence of a close binary
may explain the need for two wind components components in our spectro-astrometry, separated by about 15\,$\rm km\,s^{-1}$ at the time of
observation. It should be noted that a match with a binary is not unambiguous, and other models of extended emission along a P.A.=44\degr are also possible.   
\paragraph{T~CrA} \cite{Bailey98, Takami03} reported a binary companion, using spectro-astrometry of H$\alpha$, with a separation of $>0\farcs14$ and P.A.=278\degr. We do not detect
a companion at $4.7\,\mu$m at this P.A. and separation with an upper limit of the flux ratio of 6. Speckle imaging at K-band also does not detect the companion \citep{Ghez97, Koehler08}. 
It is possible that the companion is significantly bluer than the primary, and we do not expect it to influence our spectro-astrometry.

\end{document}